\documentclass[a4paper,10pt]{article}
\usepackage[T1]{fontenc}
\usepackage{amsmath}
\usepackage{amssymb}
\usepackage{graphicx}
\usepackage{empheq}
\usepackage{lmodern}
\usepackage{float}
\usepackage{caption}
\usepackage{algorithm}
\usepackage{multicol}
\usepackage[noend]{algpseudocode}
\usepackage{hyperref}
\usepackage{natbib}
\setcitestyle{authoryear,open={(},close={)}}
\usepackage{chngcntr}
\usepackage{placeins}

\usepackage[svgnames]{xcolor}

\usepackage{colortbl}
\usepackage{titlesec}

\usepackage{hyperref}
\makeatletter
\newcounter {subsubsubsection}[subsubsection]
\renewcommand\thesubsubsubsection{\thesubsubsection .\@alph\c@subsubsubsection}
\newcommand\subsubsubsection{\@startsection{subsubsubsection}{4}{\z@}%
                                     {-3.25ex\@plus -1ex \@minus -.2ex}%
                                     {1.5ex \@plus .2ex}%
                                     {\normalfont\normalsize\bfseries}}
\newcommand*\l@subsubsubsection{\@dottedtocline{3}{10.0em}{4.1em}}
\newcommand*{\subsubsubsectionmark}[1]{}
\makeatother

\hypersetup{
  colorlinks=true,
  citecolor=LimeGreen,
  linkcolor=Magenta,
  urlcolor=Blue
}
\def\be{\begin{eqnarray}}
\def\ee{\end{eqnarray}}

\usepackage{geometry}
\newtheorem{Theorem}{Theorem}[section]
\newtheorem{remark}[Theorem]{Remark}

\usepackage{chngpage}
\usepackage{array}
\newcolumntype{R}[1]{>{\raggedleft\arraybackslash }b{#1}}
\newcolumntype{L}[1]{>{\raggedright\arraybackslash }b{#1}}
\newcolumntype{C}[1]{>{\centering\arraybackslash }b{#1}}
\newcolumntype{P}[1]{>{\centering\arraybackslash}p{#1}}

\DeclareMathOperator{\R}{\mathbb{R}}

\usepackage{listings}
\usepackage{comment}

\def \E{\mathbb{E}}

\def\1{{\bf 1}}

\def \N{\mathbb{N}}

\usepackage{natbib}
\usepackage{bbm}

\input ./inputs/abrege
\newsavebox{\fminibox}
\newlength{\fminilength}
\newenvironment{fminipage}[1][\linewidth]
  {\setlength{\fminilength}{#1}
   \begin{lrbox}{\fminibox}\begin{minipage}{\fminilength}}
  {\end{minipage}\end{lrbox}\noindent\fbox{\usebox{\fminibox}}}

  \def\+{^\dagger}

\def\nequiv{\not\kern-.05em\equiv}
\def\egal{\kern-.5em=\kern-.5em}        
\def\propt{\kern-.2em\propto\kern-.2em} 

\def\argmin{\mathop{\mathrm{arg\,min}}} 
\def\intdouble{\int\kern-0.3em\int}
\def\inttriple{\int\kern-0.3em\int\kern-0.3em\int}

\def\rond#1{\overset{\kern-0.33em~_\circ}{#1}}
\def\rondit[#1]#2{\overset{\kern#1~_\circ}{#2}}

\input ./inputs/alphabet

\def\babs{\begin{abstract}}             \def\eabs{\end{abstract}}
\def\barr{\begin{array}}                \def\earr{\end{array}}
\def\bcc{\begin{center}}                \def\ecc{\end{center}}

\def\bdes{\begin{description}}          \def\edes{\end{description}}
\def\bdoc{\begin{document}}             \def\edoc{\end{document}}
\def\ben{\begin{enumerate}}             \def\een{\end{enumerate}}
\def\beqn{\begin{eqnarray}}             \def\eeqn{\end{eqnarray}}
\def\beqnl#1{\beqn\label{#1}}           \def\eeqnl#1{\label{#1}\eeqn}
\def\beqnx{\begin{eqnarray*}}           \def\eeqnx{\end{eqnarray*}}
\def\bseqn{\begin{subeqnarray}}         \def\eseqn{\end{subeqnarray}}
\def\beq#1\eeq{\begin{equation}#1\end{equation}}
\def\bal#1\eal{\begin{align}#1\end{align}}
\def\balx#1\ealx{\begin{align*}#1\end{align*}}
\def\beqx{$$}                           \def\eeqx{$$}
\def\bfig{\protect\begin{figure}}       \def\efig{\protect\end{figure}}
\def\bfigx{\protect\begin{figure*}}     \def\efigx{\protect\end{figure*}}
\def\bfigt{\protect\begin{figurette}}   \def\efigt{\protect\end{figurette}}
\def\bfl{\begin{flushleft}}             \def\efl{\end{flushleft}}
\def\bfr{\begin{flushright}}            \def\efr{\end{flushright}}
\def\bit{\begin{itemize}}               \def\eit{\end{itemize}}
\def\bmi{\begin{minipage}}              \def\emi{\end{minipage}}
\def\bfmi{\begin{fminipage}}            \def\efmi{\end{fminipage}}
\def\bpic{\begin{picture}}              \def\epic{\end{picture}}
\def\bqu{\begin{quote}}                 \def\equ{\end{quote}}
\def\bqun{\begin{quotation}}            \def\equn{\end{quotation}}
\def\bsl{\begin{slide}}                 \def\esl{\end{slide}}
\def\btabb{\begin{tabbing}}             \def\etabb{\end{tabbing}}
\def\btabl{\begin{table}}               \def\etabl{\end{table}}
\def\btablx{\begin{table*}}             \def\etablx{\end{table*}}
\def\btab{\begin{tabular}} 
\def\btabu{\begin{tabular}}             \def\etabu{\end{tabular}}
\def\btabx{\begin{tabular*}}            \def\etabx{\end{tabular*}}
\def\bbib{\begin{thebibliography}{999}} \def\ebib{\end{thebibliography}}
\def\bver{\begin{verbatim}}             \def\ever{\end{verbatim}}
\def\bca{\begin{cases}}                  \def\eca{\end{cases}}

\begin{document}

\title{Deep learning for discrete-time hedging in incomplete markets}
\author{Simon FECAMP  \footnote{EDF R\&D} \thanks{simon.fecamp@edf.fr}
\and
Joseph MIKAEL \footnote{EDF R\&D} \thanks{joseph.mikael@edf.fr}
\and 
Xavier WARIN \footnote{EDF R\&D \& FiME, Laboratoire de Finance des March\'es de l'Energie} \thanks{xavier.warin@edf.fr}
}
\date{February, 2019}

\maketitle

\abstract{
\textit{Several algorithms based on machine learning to solve hedging problems in incomplete markets are presented. The sources of incompleteness are illiquidity, non-tradable risk factors, discrete hedging dates and proportional transaction costs. Hedging strategies induced by the algorithms introduced in this paper are compared to classical stochastic control techniques on several payoffs using a MSE criterion.
Some of the proposed algorithms are flexible enough to deal with innovative loss criteria and P\&L distribution of the hedging strategies obtained with these new criteria are compared to P\&L distribution obtained with the classical MSE criterion. The most efficient algorithm is tested on a case with non-zero transaction costs and we show how to obtain a whole Pareto frontier in a single training phase by randomly combining the criteria of average cost and variance during the learning phase. 
\\  \vspace{5mm} }

}

\vspace{5mm}

\section{Introduction}
\label{sec:introduction}
Despite its desirable properties, the complete market assumption is destroyed as soon as we consider transactions costs, discrete time hedging dates, illiquidity, non-tradable risk factors (e.g. volume risk), ... These properties make the completeness assumption not realistic in most of the financial markets and especially when trading on commodities markets. In an incomplete market, the set of non attainable contingent claims (i.e. contingent claims that cannot be replicated by a self-financing strategy) is not empty and for these, one needs a criterion to decide how to share risks between the seller and the buyer.
The literature deals with three families of criteria: quantile hedging, utility functions and moment-based criteria.  \\ 
Quantile hedging (see \cite{follmerquantile}, \cite{bouchard2017numerical2}) aim is to construct a hedging strategy which maximizes the probability of a successful hedge given a constraint on the required
cost. Another possibility offered by quantile hedging is to set a shortfall probability $\varepsilon$ and minimize the
cost in the class of hedging strategies such that the probability of covering the claim is at
least $1-\varepsilon$. 
 \\ 
Utility-based-criteria and more precisely utility  indifference (see \cite{carmona2008indifferenc}) has the favor of academics as it sometimes allows to get analytic prices and hedging strategies. However this approach is not used by practitioners as the associated risk aversion coefficient is hard to define. \\  
The last family, that we use in this paper is based on moments of the distribution of the hedged portfolio.
The simplest moment-based-criterion is the variance criterion minimizing the variance of the hedged portfolio and the local variance of the portfolio (see \cite{schweizer1999guided} for a survey in continuous time). However quadratic criteria penalizes in the same way losses and gains. This might be seen as a drawback but this however offers the advantage of giving the same price to both buyers and sellers. \cite{gobet:hal-01761234} extends the local mean squared criterion by introducing an asymmetry in the loss function that penalizes more losses than gains. In the case of a variance criterion or a local variance criterion, continuous time hedging strategies when the assets are modeled using some Levy processes are given for example in \cite {tankov2003financial}. \\
Once the criterion has been chosen, one has to compute the trading strategy minimizing it. Specific methods must be developed to deal with the source of incompleteness (whether it is illiquidity, transaction costs, non-tradable risk factor, ...) \\ 
Limited availability of hedging products can be dealt in two ways. First, \cite{potters2003more}, \cite{gatheral2010no} or \cite{lehalle2013market} study the price impact of selling or buying an underlying on markets. The impact being greater with the exchanged volume, a seller will tend to limit the amount of volume to sell at one time. A second approach consists in assuming that in practice risk managers are aware of the liquidity constraints of the markets and try to implement strategies taking these into account.
In the case of a global variance minimization of the hedged portfolio,
\cite{2017warin} developed some algorithms based on regression to calculate the hedging strategy taking into account all of these liquidity constraints. \\ 
In the literature, transaction costs treatment comes together with discrete hedging. The pioneering work of \cite{leland1985option} proposes to use the Black-Scholes formula with a modified volatility. \cite{kabanov2009markets} gives replication bound errors to the \cite{leland1985option} model. \cite{toft1996mean} uses a mean-variance criterion to analyze the trade-off between costs and risks of discretely rebalanced option hedges in the presence of transactions costs. \\ 
In general when no closed-form-formula for the optimal hedging strategy is available we use some stochastic dynamic programming algorithms that suffer from the curse of dimensionality. To our knowledge, there exists no algorithm to define the optimal strategy with arbitrary criteria together with liquidity constraints and transaction costs and robust to high dimensions. \\ 
In this article we propose some machines learning algorithms to derive optimal hedging strategy.

 \begin{itemize}
     \item the first set of algorithms try to calculate hedging positions by solving a global risk minimization problem.
     The hedging strategies are calculated using different types of architectures. The most efficient architecture is easy to implement and can be used with liquidity constraints, general risk criteria and with transaction costs. This  algorithm is fast enough to be used in high dimensions.
    This approach is directly linked to the algorithm proposed by \cite{weinan2017deep} that uses a global optimization problem to solve semi-linear PDEs by controlling the $z$ term in the BSDE approach.
     \item the second and third algorithms are some machine learning version of the two algorithms described in \cite{2017warin} that can only be used for a variance criterion: a dynamic programming method is used and some minimization problems are solved at each time step in order to calculate the optimal hedging strategy. This approach is based on a succession of local optimizations and this kind of ideas have proved to be more effective than the global optimization approach in the resolution of non linear PDEs  \cite{hure2019some}, \cite{beck2017machine}, \cite{germain2020deep}.
 \end{itemize}
 We first describe the hedging problem, we define the price model and we present several loss functions used for the experiments. After detailing the different algorithms used, we focus on the variance criterion and compare the results obtained by the different algorithms on options involving a variable number of risk factors, be they tradable or not. We take as a reference calculations achieved on high performance computers by the StOpt library \cite{stopt} using the algorithm 2 described in \cite{2017warin}.  We then train the first algorithm with the different risk criteria already mentioned, and discuss the impact of these criteria on the distribution of the hedged portfolio. At last, we introduce transaction costs and show how to estimate a Pareto frontier by training the algorithm with random combinations of mean and variance targets.\\
 The main results of the paper are the following:
 \begin{itemize}
     \item We show that the use of deep neural network algorithms for reasonably realistic option hedging problems (discrete-time, unhedgeable factors, limited liquidity, transaction costs, general risk criteria) is possible.
     \item The comparison of global and local neural network architectures is achieved for a variance criterion. At the opposite of PDE resolution, the global approach appears to be more effective than local minimization approach as it is far less costly and often more accurate.
     The local minimization approach is often trapped in local minima and near optimal solutions are obtained by running the algorithm many times. This difference in behavior compared to semi-linear PDE resolution is certainly related to the fact that in PDE resolution the direct process used as state is not controlled.
     In our case, part of the state is controlled, and we need to sample the state randomly using an a priori law. This seems to the key point explaining
     the superiority of the global approach.
    The references, obtained by dynamic programming and regressions, can only be calculated in small dimension, but we expect the results to be still very good in higher dimension.
    \item Using the global algorithm, we are able to solve efficiently some hedging problems that are out of reach by classical dynamic programming methods due the structure of the risk function used. For example, the computation of a Pareto efficient frontier is now possible.
 \end{itemize}
 
\section{Problem description}
In the numerical tests, we retain the price modelling used in \cite{2017warin}. A short description is done in Section \ref{sec:notations} and we refer to the original paper for further details.
\subsection{Risk factors modelling}
\label{sec:notations}
We are given a financial market operating in continuous time: we begin with a probability space $(\Omega, \Fc, \Pbb)$, a time horizon $0<T<\infty$ and a filtration $\Fc = (\Fc_t),$ ${0\leq t\leq  T}$ representing the information available at time $t$. We consider $d + 1 $ assets $\hat{F}^0, \ldots, \hat{F}^d$ available for trade. For the sake of simplicity, we suppose a zero interest rate and we assume that there exists a risk free asset $\hat{F }^0$ having a strictly positive price. We then use $\hat{F}^0$ as numeraire and immediately pass to quantities discounted with $\hat{F}^0$. We denote $F^i = \hat{F}^i/\hat{F}^0$, $i=1, ...d$ the thus discounted quantities and $F$ the vector having the $(F^i)_{i=1..d}$ as coordinates. We consider another non tradable risk factor (the volume risk) denoted $\cal V$. \\ 
The evolution of the $(F^i)_{i = 1..d}$ and of ${\cal V}$ are respectively described by a diffusion process having values in $\Rbb^d$ and in $\Rbb$. \\ 
More precisely, the volume risk ${\cal V}_t$ is stochastic and follows for $t\geq u\geq 0$ the dynamic: 
\begin{eqnarray}
\label{eq:volume}
{\cal V}_t & = & \hat{\cal V}_t + \left({\cal V}_u-\hat{\cal V}_u\right)e^{-a_{\cal V}(t-u)} + \int_u^t \sigma_{\cal V} e^{-a_{\cal V} (t - s)}dW_s^{\cal V}
\end{eqnarray}
where $a_{\cal V}$ is the mean reverting coefficient, $\sigma_{\cal V}\geq 0$ the volatility,  and $W_t^{\cal V}$ is a Brownian motion on $(\Omega, \Fc, \Pbb)$. $\hat{\cal V}_u$ is the average load seen on the previous years  at the given date $u\geq 0$. This mean reverting model is generally used to represent the load dynamic of some electricity contracts. 
We suppose that, for $i = 1, \ldots, d$, the prices are martingales and  follow the dynamic: 
\begin{eqnarray}
F^i_t& =& F^i_0 e^{-(\sigma_{i, E}) ^2 \frac{e^{-2 a_{i, E}(T-t)} - e^{-2 a_{i, E} T}}{4 a_{i, E}} + e^{-a_{i, E}(T-t) }\hat{W}_t^{i, E} }, \nonumber\\
\hat{W}_t^{i, E}& =& \sigma_{i, E}\int_0^t e^{-a_{i, E} (t-s)} dW_s^i \label{eq:futurediff}
\end{eqnarray}
where $F^i_t$  represents the forward price seen at time $t$ for a delivery at date $T$ which is given once for all and will correspond to the maturity of the considered contracts, $a_{i, E}$ the mean reverting parameter for risk factor $i$, $\sigma_{i,E}$ the volatility parameter for risk factor $i$ and $W_s^i$ a Brownian motion on $(\Omega, \Fc, \Pbb)$ so that the $W_t^i$ are correlated and also correlate with $W_t^{\cal V}$ . We will denote $S_t$ the vector $(F^1_t, \ldots, F^d_t, {\cal V}_t)$. 
\subsection{Hedging problem}
\label{sec:hedgingproblem}
We consider the hedging problem of a contingent claim paying $g(S_T)$ at time $T$ where $S_T$ denotes the contingent claim underlying vector. Without loss of generalities, in the following we consider ourselves as the derivative seller. 
We consider a finite set of hedging dates $t_0<t_1<\ldots<{t_{N-1}}<\ldots < {t_{N}} = T.$ The discrete hedging dates bring the first source of incompleteness.
At each date, each of the discounted assets $F^i$ can only be bought and sold at a finite quantity $l^i$ giving a second source of incompleteness. The volume risk ${\cal V}_ t$ cannot be traded and  is the third source of incompleteness. 
A self-financing portfolio is a $d$-dimensional $(\Fc_t)$-adapted process $\Delta_t$.
Its terminal value at time $T$ is denoted $X_T^\Delta$ and satisfies:

\begin{eqnarray*}
    X_T^\Delta &=& p +\sum_{i=1}^d \sum_{j=0}^{N-1} \Delta^i_{t_j}(F^i_{t_{j+1}} -  F^i_{t_{j}}),
\end{eqnarray*}
where $p$ will be referred to as the premium. 
Between two time steps, the change in $\Delta^i$, corresponding to the buy or sell command $C^i_{j+1}: =  \Delta^i_{t_{j+1}} - \Delta^i_{t_{j}}$  should not exceed in absolute value the liquidity $l^i$ so that: 
\begin{eqnarray*}
  | \Delta^i_{t_0}| \leq l^i , &   |C^i_j| \leq l^i,  \quad j=1, \dots ,N-1, i =1,\dots,d.
\end{eqnarray*}
Given a loss function $L$, and denoting $Y_T = X^\Delta_T- g(S_T))$, we search for a strategy verifying:
\begin{equation}
 (p^{Opt}, \Delta^{Opt}) = Argmin_{p,\Delta}{L(X^\Delta_T- g(S_T)) } =Argmin_{p,\Delta}{L(Y_T) } \label{eq:description}.
\end{equation}

We will focus on the following loss functions: 
\bit
\item \textbf{Mean Squared error} defined by 
\begin{equation}
\label{eq:riskL2}
  L(Y) = \Ebb\left[Y^2 \right].
\end{equation}
It has been intensively studied for example in \cite{schweizer1999guided}. It has the drawback of penalizing losses and gain the same way. This also can be seen as an advantage as it gives the same value and strategy for the buyer and for the seller. 
\item \textbf{Asymmetrical loss} defined by:  
\begin{equation}
  \label{eq:riskAsym}  
  L^\alpha(Y )= \Ebb\left[(1+ \alpha) Y^2\1_{ Y\leq 0} + Y^2 \1_{Y\geq 0} \right].
\end{equation}
When $\alpha > 0$ (resp. $0 < \alpha$)  , the losses  (resp. gains) are penalized.
It will be referred to as the asymmetrical loss. It has been studied for example in \cite{gobet:hal-01761234}.

\item \textbf{Loss Moment 2/Moment 4 function} defined by:
\begin{equation}
    \label{eq:variancemoment}
    L^{\alpha}(Y)= \Ebb\left[Y^2 \1_{ Y\geq 0}\right] + \alpha \Ebb\left[Y^4 \1_{Y\leq 0} \right], \alpha \geq 1.
\end{equation}
This criteria is designed to penalize heavy tail on the loss side. 
\eit 

\section{Some classical neural networks and  stochastic gradient algorithms}
Deep neural networks are state-of-the-art tools for approximating functions (see \cite{liang2017deepnn}).
This section present two classical network used in machine learning. The first one is well known in the stochastic optimization community as it is used for example in  \cite{weinan2017deep}, \cite{hure2019some}, \cite{beck2017machine}, \cite{germain2020deep} to solve some non linear PDE problem. The second gives the possibility the treat some non Markovian problems  at a given date $t$  as the whole past of the trajectories is kept in  memory.
As the neural approximation of a function is highly non linear, it always lead to a non convex optimization problem that need some specific resolution algorithm that we detail.
\subsection{Feedforward neural network as function approximators}
We suppose in this section that the input is in dimension $d_0$ (the state variable $x$) and the output is in dimension $d_1$ (the number of  value functions to estimate).
The network is characterized by a number of layers 
 $L+1$ $\in$ $\N\setminus\{1,2\}$  with $m_\ell$, $\ell$ $=$ $0,\ldots,L$, the number of neurons (units or nodes) on each layer: the first layer is the input layer with $m_0$ $=$ $d_0$, the last layer is the output layer with $m_L$ $=$ $d_1$, and the $L-1$ layers between are called hidden layers, where we choose for simplicity the same  dimension $m_\ell$ $=$ $m$, $\ell$ $=$ $1,\ldots,L-1$.
 A  feedforward neural network is a  function from  $\R^{d_0}$ to $\R^{d_1}$ defined as the composition
\begin{align} \label{defNN}
x \in \R^{d_0}  & \longmapsto  \; A_L \circ  \varrho \circ A_{L - 1} \circ \ldots \circ \varrho \circ A_1(x) \; \in \; \R. 
\end{align}
Here $A_\ell$, $\ell$ $=$ $1,\ldots,L$ are affine transformations: $A_1$ maps from $\R^{d_0}$ to $\R^m$, $A_2,\ldots,A_{L-1}$ map from $\R^m$ to $\R^m$, and $A_L$ maps from $\R^m$ to 
$\R^{d_1}$, represented by 
\begin{align*}
A_\ell (x) &= \; \Wc_\ell x + \beta_\ell,
\end{align*}
for a matrix $\Wc_\ell$ called weight, and a vector $\beta_\ell$ called  bias term,  $\varrho$ $:$ $\R$ $\rightarrow$ $\R$ is a nonlinear function, called activation function, and applied 
component-wise on the outputs of $A_\ell$, i.e., $\varrho(x_1,\ldots,x_m)$ $=$ $(\varrho(x_1),\ldots,\varrho(x_m))$. Standard examples of activation functions are the sigmoid, the ReLu, the Elu, $\tanh$. 
All these matrices $\Wc_\ell$ and vectors $\beta_\ell$, $\ell$ $=$ 
$1,\ldots,L$,  are the parameters of the neural network, and can be identified with  an element $\theta$ $\in$ $\R^{N_m}$, where $N_m$ $=$ 
$\sum_{\ell=0}^{L-1} m_\ell (1+m_{\ell+1})$ $=$ $d_0(1+m)+m(1+m)(L-2)+m(1+d_1)$ is the number of parameters.  
The universal approximation theorem of Hornick et al. \cite{hor90universal}  states that set all feedforward approximators making $m$ vary is dense in $L^2(\nu)$ for any finite measure $\nu$ on $\R^d$, $d>0$ whenever $\varrho$ is continuous and non-constant.  \\ 
Assuming the optimal control of Equation \eqref{eq:description} is sufficiently smooth, from the universal approximation theorem we do know that the control can be approached with a feedforward neural network having sufficient depth and width. The latter theorem does not tell what are the minimal depth and width so that empirical studies have to be done to know what is the best architecture. The universal approximation theorem does not tell neither how to optimize the neural networks weights but it appears that a stochastic gradient descent shows good results in many cases.

\subsection{Recurrent and classical LSTM neural networks as time-dependent-function approximator}
Recurrent neural networks (RNNs) are dynamical systems that make efficient the use of temporal information in the input sequence. For RNNs the input is a times series and in this paper, the output is composed of two vectors: a memory state $M_t$ and an output state $C_t$. At each time step $t$, $M_{t-1}$ and $C_{t-1}$ are given together with the time series to a recurrent cell i.e. a neural network whose weights are shared across all time steps (see Figure \ref{fig:recurrent}). 
Long short term memory cells (\cite{article:lstm}) are powerful for capturing long-range dependence of the data. They are designed to avoid some vanishing gradients effect that basic RNN suffers. In an LSTM cell, structures called gates regulates the flow of information contained in the memory state $M_t$ by adding or removing information to the state. Gates are composed out of a sigmoid neural network layer and a pointwise multiplication operation. Mathematically, the rules inside the $t$-th cell follows: 
\begin{eqnarray}
\Gamma^f_t &= &\sigma(A_f S_t + U_f C_{t-1} + b_f)  \nonumber\\
\Gamma^i_t  &= &\sigma(A_i S_t + U_i C_{t-1} + b_i) \nonumber \\  
\Gamma^o_t  &=& \sigma(A_o S_t + U_o C_{t-1}+ b_o) \nonumber\\
M_t & = & \Gamma^f_t \odot M_{t-1} + \Gamma^i_t \odot tanh(A_M S_t + U_M C_{t-1} + b_M), M_0 = 0 \nonumber\\
C_t & =& \Gamma^o_t \odot tanh(M_t), C_0 = 0 \label{eq:lstm}
\end{eqnarray}
where $\odot$ is the Hadamard product, $\sigma$ is the sigmoid activation function $\left(\sigma(x) = \frac{1}{1+e^{-x}}\right)$, $A_{\bullet} \in \mathbb{R}^{h \times d}$, $U_{\bullet}^{h\times h}$, $b_\bullet \in\mathbb{R}^h$, $h$ being the cell state size. $\Gamma^f_t$ represents the forget gate. It  decides what information needs to be deleted from the memory state. This decision is made by a sigmoid layer called the ``forget gate layer''. It outputs a number between 0 and 1 and multiply it to each number in the memory state $M_{t−1}$. $\Gamma^i_t$ is the input gate evaluating what new information needs to be stored in the memory state. The output gate layer $\Gamma^o_t$ decides what parts of the memory state needs to be outputted. It is based on filtered version of the memory state. The weight matrices and bias vector ($A_{\bullet}, U_{\bullet}, b_{\bullet}$) are shared through all time steps and are learned during the training process.  The output $C_t$ is used as an approximation of the unknown function. We still note $\theta$ the set of parameters used for the LSTM representation \eqref{eq:lstm}.

\subsection{General optimisation algorithm}
As the  use of neural network leads to highly non convex and non linear optimisation problems, we use  a mini-batch stochastic gradient descent for calculate the $\theta$ parameters. Adaptive Moment Estimation (Adam) \cite{kingma2014adam} is a method that computes adaptive learning rates for each parameter. In addition to storing an exponentially decaying average of past squared gradients $v_t$ like AdaDelta \cite{zeiler2012adadelta} and RMSprop (\cite{tieleman2012lecture} ) Adam also keeps an exponentially decaying average of past gradients $m_t$ similar to momentum.

\begin{algorithm}[H]
\caption{\label{algoGlobal}Forward resolution of global algorithms}
\begin{algorithmic}[1]
 \State $\alpha$ : Stepsize
 \State $\beta_1$, $\beta_2\in[0,1],$ Exponential decay rates for the moment estimates,
 \State $N_{iter}$ number of iterations
 \State $N_{batch}$, the number of simulations at each gradient descent iteration (batch size). 
\State $\theta_0$ randomly chosen 
\State $m_0 \leftarrow 0 $
\State $v_0 \leftarrow 0$ 
\State $t \leftarrow 0 $
\For{$t=0\ldots N_{Iter}$}
\State $S_u \leftarrow N_{batch}$ samples simulations of $S_u$, $u =t_0, ...,t_{N-1}, T$
\State $t \leftarrow t +1 $
\State $g_t = \nabla_\theta L(\mathbb{NN}^{\theta_{t-1}}(S_u) - g(S_T)) $ (get gradient w.r.t objective function) 
\State  $m_t \leftarrow m_{t-1} + (1-\beta_1) . g_t$ (update biased first moment estimate) 
\State $v_t \leftarrow \beta_2 v_{t-1} + (1-\beta_2) g_t ^2$ (update biased second raw moment estimate)
\State $\hat{m}_t  \leftarrow \frac{m_t}{1-\beta_1^t}$ (computes bias-corrected first moment estimate ($\beta_1^t$ stands for $\beta_1$ to the power of $t$)) 
\State $\hat{v}_t  \leftarrow \frac{v_t}{1-\beta_2^t}$ (computes bias-corrected second raw moment estimate ($\beta_2^t$ stands for $\beta_1$ to the power of $t$))
\State $\theta_t \leftarrow \theta_{t-1} - \alpha \hat{m}_t /(\sqrt{\hat{v}_t}+\epsilon )$ (update parameters)

\EndFor
\end{algorithmic}
\end{algorithm}

\section{Optimal network for the global hedging problem}
In this section, we compare the  use of two feed forward  networks to the LSTM network and an extension of the LSTM network that we propose.
We first explain how to use the previously proposed feed forward network in the context of hedging a call option in the Black Scholes model. We then  detail our LSTM extension and
 compare the results obtained on the hedging problem without any hedging constraints. We show that the modified LSTM network give the best results.
At last we explain how to adapt  our modified LSTM network to deal with liquidity constraints. 
In the following,  $\tilde{S}_t = \frac{S_t -\E[S_t]}{\sqrt{\E[(S_t-\E(S_t))^2]}}$  denotes a normalized version of $S_t$.

\subsection{Feedforward neural networks architectures on the  global hedging problem}
\begin{figure}
    \centering
    \includegraphics[width=0.7\textwidth]{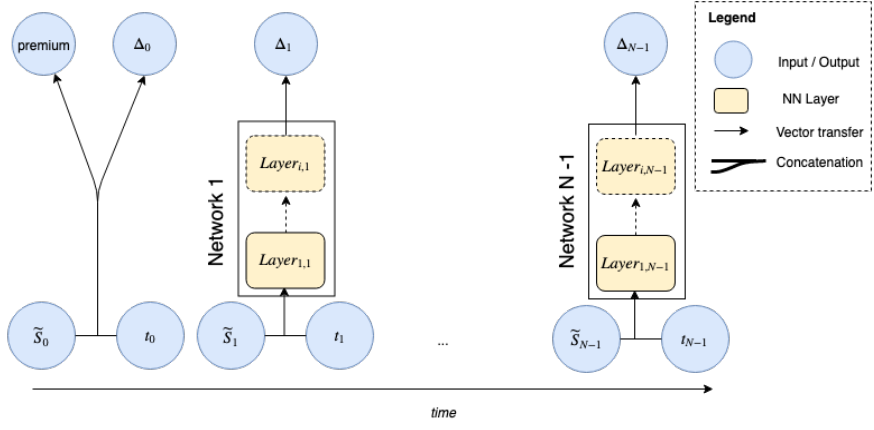}
    \caption{Feedforward basic architecture: $N-1$ feedforward networks for the $N-1$ time steps}
    \label{fig:deltafeedforward}
\end{figure}

\begin{figure}
    \centering
    \includegraphics[width=0.7\textwidth]{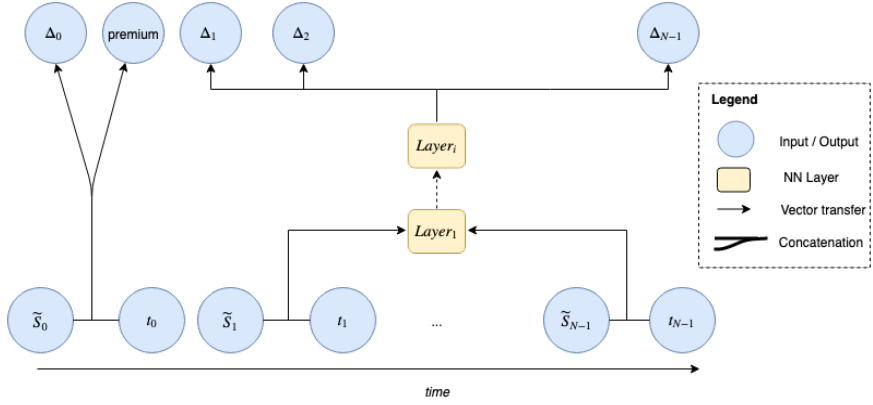}
    \caption{Feedforward merged architecture: a time dimension is added to the input features but the feedforward bias and weights networks are shared within all the timesteps}
    \label{fig:deltafeedforwardmerged}
\end{figure}

A possible approach to solve the hedging problem described in Section \ref{sec:hedgingproblem} consists in training $N$ different feedforward neural networks (one per time steps) as done in \cite{Han8505} for the PDE case and as illustrated in Figure \ref{fig:deltafeedforward} and described in Section \ref{sec:feedforwardcontrol}. This architecture (denominated feedforward control in the following) generates a possibly high number of weights and bias to be estimated ($N * depth * width$). Another possibility is to train one single feedforward neural network fed with the prices and the time to maturity as proposed in \cite{Chan-Wai-Nam2019} and as illustrated in Figure \ref{fig:deltafeedforwardmerged} and described in Section \ref{sec:feedfowardmergecontrol}. This architecture is referred to as feedforward merged control in what follows. 

\subsubsection{Feedforward control structure}
\label{sec:feedforwardcontrol}
In the feedforward control network, $N-1$ networks are fed successively with $(\tilde{S}_{t_i})_{i =1...N-1}$. The feedforwards networks are parameterized by  $\theta$ (the bias and weights to be estimated). The $i$-th feedforward neural network provides a $d$ dimensional control $\Delta_{t_i}(\tilde{S}_{t_i}, \theta)$. The first control $\Delta_{t_0}(\tilde{S}_{t_0},\theta)$ and the premium $p(\theta)$ are trainable variables. 
The final payoff is given by:
\begin{eqnarray*}
    X_T(\theta) &=& p(\theta)+\sum_{i=1}^d \sum_{j=0}^{N-1} \Delta^i_{t_j}(\tilde S_{t_j}, \theta) (F^i_{t_{j+1}} -  F^i_{t_{j}}).
\end{eqnarray*}
and the problem \eqref{eq:description} leads to  the following optimization problem:\\
\begin{equation}
 \theta^* =  Argmin_{\theta}{L(X_T(\theta)- g(S_T)) }\label{eq:feedforwardproblem}.
\end{equation}

\subsubsection{Feedforward merged control structure}
\label{sec:feedfowardmergecontrol}
In the feedforward merged control structure a single neural network is fed successively with  $(\tilde{S}_{t_i})_{i = 1...N-1}$. For each pair $(t_i, \tilde S_{t_i})$ the network provides a control $\Delta(t_i, \tilde S_{t_i}, \theta)$  where $\theta$ represents the bias and weights to be estimated. Again, the first control $\Delta(t_0, \tilde{S}_{t_0},\theta)$ and the premium $p(\theta)$ are trainable variables. 
The final payoff is given by:
\begin{eqnarray*}
    X_T(\theta) &=& p(\theta)+\sum_{i=1}^d \sum_{j=0}^{N-1} \Delta^i(t_j , \tilde S_{t_j}, \theta) (F^i_{t_{j+1}} -  F^i_{t_{j}}).
\end{eqnarray*}
The problem \eqref{eq:description} leads to  the following optimization problem:\\
\begin{equation}
 \theta^* =  Argmin_{\theta}{L(X_T(\theta)- g(S_T)) }\label{eq:feedforwardmergedproblem}.
\end{equation}

 \subsection{Recurrent networks}
 
\begin{figure}
\begin{center}
\includegraphics[width = 0.9\textwidth]{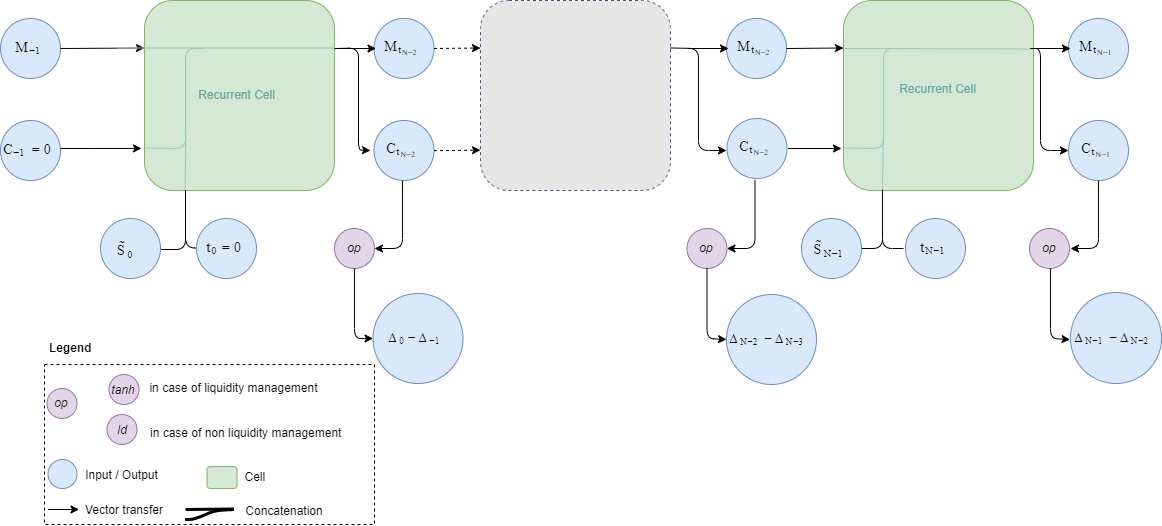}
\caption{Recurrent architecture. The cell can be a LSTM cell for instance. }
\label{fig:recurrent}
\end{center}
\end{figure}

\begin{figure}
\begin{center}
\includegraphics[width = 0.8\textwidth]{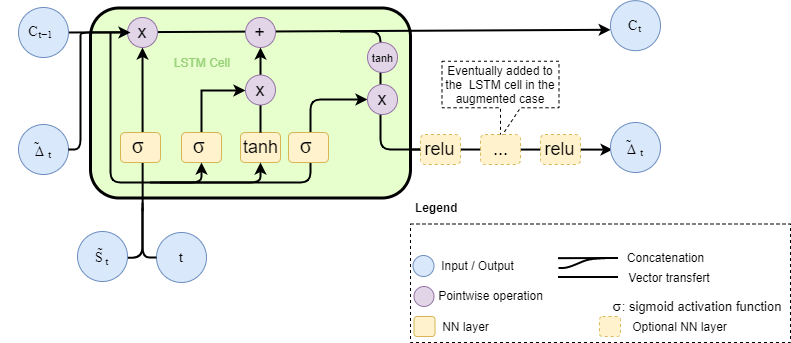}
\caption{LSTM Cell possibly combined with a feedforward network (Figure inspired by \cite{colah2015understanding})}
\label{fig:lstm}
\end{center}
\end{figure}

The hedging problem sequential nature makes relevant the use of recurrent neural networks (RNN). This kind of networks is used for example in \cite{Chan-Wai-Nam2019} for the PDE numerical resolution problem. As mentioned in \cite{chung2014empirical}, among all RNN architectures, LSTM neural networks (see \cite{article:lstm}) present several advantages among which the convergence speed and the memory management. It would allow for example the management of non Markovian underlying models. This architecture will be referred to as classical LSTM in the following. 
As more layers may represent more complex functions of the inputs, we propose to test whether the addition of a feed forward network to the LSTM cell output as shown in Figure \ref{fig:lstm} helps the algorithm to converge. This composition of LSTM cell and feed forward network is referred to as augmented LSTM cell in the following.\\
The recurrent cell is fed with $\tilde{S}_t$. Its recursive calls on a sequence of inputs provides a sequence of underlying positions changes (see Figure \ref{fig:recurrent}).
At each date $t_j$, the recurrent cell produces a $d$-dimensional output depending on historical events and controls  $\hat{C}_j(\theta, (\tilde{S}_{t_s})_{s\leq j}, (\Delta_{t_s})_{s\leq j} ))$ (denoted simply $C_j$ in Figure \ref{fig:lstm}) that is not bounded. 
The strategy $\Delta$'s are calculated for $j= 0, \dots, N-1; i=1, \dots, d$
\begin{eqnarray}
\Delta^i (t_j, (\tilde S_{t_i})_{i\leq j}, \theta) & =&   \sum_{k=0}^j  \hat{C}_k^i( (\tilde{S}_{t_s})_{s\leq j}, \Delta(t_s,(\tilde S_{t_s})_{s\leq j}, \theta) ). \label{eq:deltalstm1}
\end{eqnarray} 

The final payoff is then given by:
\begin{eqnarray*}
    X_T(\theta) &=& p(\theta)+\sum_{i=1}^d \sum_{j=0}^{N-1} \Delta^i(t_j, (\tilde S_{t_k})_{k\leq j}, \theta)(F^i_{t_{j+1}} -  F^i_{t_{j}}).
\end{eqnarray*}
and the problem \eqref{eq:description} leads  the following optimization problem:\\
\begin{equation}
 \theta^* =  Argmin_{\theta}{L(X_T(\theta)- g(S_T)) }\label{eq:descriptionDis}.
\end{equation}

\subsection{Neural networks extra-parameters}
\label{sec:parametrization}
The neural networks results depend on some extra parameters listed hereafter. Unless otherwise specified, these parameters are shared for all the test cases.
\begin{itemize}
\item The \textbf{batch size}, the number of simulations we give at each iteration of the Adam optimizer is equal to 50. 
\item The Adam initial \textbf{learning rate} is equal to 0.001 (default parameter). 
\item The \textbf{number of units in LSTM cells} (dimension of $M_t$) in the LSTM cell is equal to 50.
\item We use 3 ReLU layers and densities of the feedforward part of 10 for the augmented LSTM cell. 
\item We use \textbf{batch normalization} of the data before they are given to the neural networks. The mean and variance used for the normalization are computed once for all over a subset of 100 000 simulations.
\item Unless otherwise specified the \textbf{number of iterations} in the gradient descent algorithm is equal to 20 000. Every 1 000 iterations, we keep the neural network state if it gives a better loss on the test set than previously. 
\end{itemize}
In the following, tests are done using TensorFlow (\cite{2015tensorflow}).

\subsubsection{Numerical comparison of global neural-network architectures}
\label{sec:comparisonfflstm}

Table \ref{tab:bscomparison} compares the Mean squared hedging error of Equation \eqref{eq:riskL2} obtained with these two architectures and with the augmented LSTM architecture  for a Black-Scholes call option (with trend $\mu$ and volatility $\sigma$) with no liquidity constraints. Results obtained with the Black-Scholes $\Delta$ are also shown. After 20 000 iterations, the results obtained with the augmented LSTM architecture are better than with the two feedforward networks.  Of course, the Black-Scholes $\Delta$ is in this case  \textit{almost} complete market setting is unbeatable and the Black-Scholes error would be 0 in a continuous time implementation. However we can see that  the replication error given by the augmented LSTM is relatively low comparatively to Black-Scholes.

\begin{table}[!ht]
\centering
\begin{tabular}{|l|l|}
\hline
\textbf{}  & \textbf{Mean Squared error}  \\ \hline
Black-Scholes $\Delta (N(d_1))$ & 1.61e-05 \\ \hline \hline 
Feedforward delta [10, 10, 10]& 1.32e-04 \\  \hline 
Feedforward delta [10, 15, 30]& 1.31e-04
 \\  \hline 
Feedforward merged [10, 10, 10] & 1.37e-04 \\ \hline 
Feedforward merged [10, 15, 30] & 
1.30e-04\\ \hline 
Augmented LSTM 50 units [10, 10, 10]  &  1.73e-05   \\ \hline 
\end{tabular}
\caption{Mean Squared error on a Black-Scholes call option with different neural network architectures.  Layer sizes are denoted with a list (e.g. [10, 15, 20] means three hidden layers of sizes respectively 10, 15 and 20). Parameters: $S_0 = K = 1, \Delta t = 1/365, T = 1/12 \hbox{ years}, \mu = 0, \sigma = 0.2).$ The number of iterations is set to 20 000. Activation functions for the feedforward networks are ReLu functions. }
\label{tab:bscomparison}
\end{table}
In Table \ref{tab:basicvsaugmentedlstm}, we show the Mean Squared error of Equation \eqref{eq:riskL2} loss derived from a classical LSTM cell on a liquidity-constraints-free vanilla call option and on a 2 market spreads call option (having payoff $(S_T^1 -S^2_T-K)^+$).  We compare this loss to the loss derived from the augmented LSTM cell. We can see that for the more complex payoff represented here by a 2 markets spread, the augmented LSTM cell gives slightly better results.


\begin{table}[!ht]
\centering
\begin{tabular}{|l|l|l|}
\hline
\textbf{}  & Black Scholes call option & 2 markets spread \\ \hline
Classical LSTM Cell & 5.73e-05
  & 3.64e-04 
 \\ \hline  
Augmented LSTM Cell & 3.97-05 & 1.11E-04  \\ \hline 
\end{tabular}
\caption{Mean Square comparison between, different classical and augmented LSTM architectures. Parameters: Call option: $T= 3/12, \Delta t = 1/360, S_0^1, \mu = 0.02, \sigma^1 = 0.3$ - 2 Markets spread option ($S_0^1 = 1., S_0^2 =0.5, K= 0.5, \sigma^1 = \sigma^2 = 0.3, \mu^1 = \mu^2 =  2\%, corr(W^1, W^2) = 0.2$).}
\label{tab:basicvsaugmentedlstm}
\end{table}
\subsection{Adaptation of the recurrent architecture to deal with liquidity constraints}
As  the control in the case of liquidity constraints  is bounded, the output of the network has  to be transformed  and we propose to use a $\tanh$ activation function as follows:
\begin{eqnarray}
\Delta^i (t_j, (\tilde S_{t_i})_{i\leq j}, \theta) & =&   l^i \sum_{k=0}^j  \tanh(\hat{C}_k^i( (\tilde{S}_{t_s})_{s\leq j}, \Delta(t_s,(\tilde S_{t_s})_{s\leq j}, \theta) )).
\end{eqnarray}
By the way, the control difference between two time steps belongs to $[-l_i, l_i]$.

\section{Local algorithms for the hedging problem with constraints}
The two other algorithms used to solve the hedging problem with constraints are local algorithms based on a dynamic programming principle proposed in \cite{2017warin}. The objective function to minimize is given by equation \eqref{eq:description}, \eqref{eq:riskL2}, so corresponds to a global variance hedging problem.
In the original article the author uses some grids for the discretization of the asset level and some regressions to calculate conditional expectations. As previously stated, theses two algorithms are only available to optimize variance problems.\\
It can be noticed the two local machine learning algorithms proposed can be related to  some recent works in \cite{hure2018deep,bachouch2018deep} and \cite{hure2019some}.\\
We introduce the spaces for $\tilde \Delta $ in $\R^d$ 
\begin{flalign*}
W_i( \tilde \Delta) = & \{ (V, \Delta) \in   \R \times \R^d,  \Fc_{t_i}\mbox{-adapted with } | \Delta^k - \tilde \Delta^k| \le  l^k , \mbox{ for } k=1, \dots, d\},\\
\Theta_i(\tilde \Delta) = & \{(\Delta_i,\ldots, \Delta_{N-1})  ,  \mbox{ where for } j \ge i , \Delta_j \mbox{ are } \R^d \mbox{ valued } \\
 & \Fc_{t_j}\mbox{-adapted with }  | \Delta_i^k - \tilde \Delta^k|  \le l^k   , |\Delta_{j+1}^k - \Delta_j^k|  \le l^k \mbox{ for }   i \le j < N-1 , k=1, \dots, d \}\\
\hat W_i(\tilde \Delta) = & \{ (V, \Delta)  \mbox{ where } V \mbox{ is } \R \mbox{ valued, } \Fc_{t_i}\mbox{-adapted ,  }  \Delta \in  \Theta_i(\tilde \Delta)\}.
\end{flalign*}
As shown in proposition 3.1 in \cite{2017warin}, the problem \eqref{eq:description}, \eqref{eq:riskL2}
can be written as
 \begin{flalign}
\label{argminTer}
(\hat p, \hat \Delta) = & \argmin_{ p \in \R , \Delta \in  \Theta_0(0)}    \sum_{i=2}^{N} \E\left[ \left(V_{i} - \sum_{k=1}^d \Delta^k_{i-1}  (F_{t_i}^k-F_{t_{i-1}}^k) - V_{i-1}\right)^2\right] + \nonumber  \\
& \E\left[\left(V_{1} - \sum_{k=1}^d \Delta^k_{0}  (F_{t_1}^k-F_{t_0}^k)- p\right)^2\right],
\end{flalign} 
where the $V_i$ satisfies:
\begin{flalign}
\label{VDef}
V_{N} =&  g(S_T), \nonumber \\
V_{i} =& \E\left[ g(S_T) - \sum_{k=1}^d \sum_{j=i}^{N-1} \Delta^k_{j}(F^k_{t_{j+1}} -  F^k_{t_{j}}) \quad |\Fc_{t_i}\right], \forall i = 1,\ldots,N-1,
\end{flalign}

\subsection{First local algorithm}
Equation \eqref{argminTer} gives a dynamic programming algorithm: introducing the optimal residual $R$ at date $t_i$, for  current state $S_{t_i}$ and having in  portfolio an investment in $\Delta_{i-1}$ assets:
\begin{align}
\label{eq:RArgmin}
R(t_i, S_{t_i}, \Delta_{i-1}) =  \min_{ (V , \Delta) \in  \hat W_i(\Delta_{i-1}) }\E\left[ \big (g(S_T) - \sum_{k=1}^d \sum_{j=i}^{N-1} \Delta^k_{j}(F^k_{t_{j+1}} -  F^k_{t_{j}})  -V  \big)^2 \quad |\Fc_{t_i}\right],
\end{align}
then equation \eqref{argminTer} gives
\begin{align}
\label{eq:progdyn0}
R(t_i, S_{i}, \Delta_{i-1}) =  & \min_{ (V , \Delta) \in  W_i(\Delta_{i-1}) }\E\left[ \big (  \tilde V -  \sum_{k=1}^d \Delta^k_{i}  ( F^k_{t_{i+1}}-F^k_{t_i})   -  V \big)^2  + R(t_{i+1}, S_{t_{i+1}},\Delta_i  ) |\Fc_{t_i}\right]
\end{align}
where $\tilde V$  is the first component of the argmin in equation \eqref{eq:RArgmin} calculating $R(t_{i+1}, S_{t_{i+1}},\Delta_i  )$.\\
In the special case where the prices are martingale the $(\tilde V, V)$ in \eqref{eq:progdyn0} are independent of the hedging strategy and  given by $(\E[g(S_T)|\Fc_{t_{i+1}}],\E[g(S_T)|\Fc_{t_{i}}]) $.
Then
\begin{align*}
    R(t_0, S_{t_0}, 0) =  & \min_{ \Delta \in \Theta_0(0)} \E\left[ \sum_{i=0}^{N-1} (  \E[g(S_T)|\Fc_{t_{i+1}}] -  \sum_{k=1}^d \Delta^k_{i}  (F^k_{t_{i+1}}-F^k_{t_i})   -  \E[g(S_T)|\Fc_{t_{i}}] \big)^2 \right]
\end{align*}
and only the hedging strategy is left to calculate by solving the classical local min variance problem leading to minimize at each time step:
\begin{align}
\label{eq:progdyn}
 \min_{  \Delta \in \R^d }\E\left[ \big (  \tilde V -  \sum_{k=1}^d \Delta^k  ( F^k_{t_{i+1}}-F^k_{t_i})   -  V \big)^2  |\Fc_{t_i}\right].
\end{align}
Our goal is then to use a neural network to calculate the $V_i$ functions (so only calculate a conditional expectation)  and the optimal control $\Delta_{i}$  both as functions of $S_{t_i}$ at each date $t_i$  by minimizing \eqref{eq:progdyn} at each time step by a backward recursion.\\
Unlike $V$, the delta have bounded values due to liquidity constraints and 
 $\Delta_{j} \in [\underline{\Delta}_j, \overline{\Delta}_j]$ where the minimal constraints $\underline{\Delta}_j$ and maximal constraints $\overline{\Delta}_j$ are in $\R^d$. \\
Normalizing the position in hedging products, we introduce $ \hat \Delta_j = \psi_j(\Delta_{j}) := \frac{ \Delta_{j} - \underline{\Delta}_j}{\overline{\Delta}_j-\underline{\Delta}_j}$ such that $ \hat \Delta_j \in [0,1]$. \\
At each time step a Feed Forward Neural Network is used to parametrize the portfolio value and the normalized command  as a function of the normalized uncertainties and the positions: $\big(\hat V_j( \theta_j; \hat S_{t_j}, \hat \Delta_j), \hat C(\theta_j; \hat S_{t_j}, \hat \Delta_j)\big)$
The first algorithm \ref{algoMeanVar1}  solves in a backward recursion \eqref{eq:progdyn}. 
Then at each time step, the resolution of equation \eqref{eq:mainAlg1} is achieved by using a machine learning approach where each function depends on some normalized variables to ease convergence of the method.
The resolution of equation \eqref{eq:mainAlg1} is achieved by using a classical stochastic gradient descent.
\begin{remark}
We create a single network for $\hat V_j$ and $\hat \Delta_{j}$ letting $\hat V_j$ depend on $\hat \Delta_{t_{j-1}}$  the  hedging position at the previous date.
In this martingale case it would have been possible to create two networks , the second being used to represent $V$ as a function of $\hat S$ only. 
\end{remark}
\begin{remark}
The position $x$ in the hedging position (normalized in $[0,1]^d$) is sampled uniformly in the algorithm. The $ \hat S_{t_j}$ are sampled according to their own empirical laws and the $ \hat S_{t_{j+1}}$ are sampled conditionally to the $ \hat S_{t_j}$.  
\end{remark}
 \begin{remark}
 The output of the Neural network has unbounded values. In order to satisfy the constraints on the hedging positions, a  $\tanh$ transformation of the output of the neural network $\hat C(\theta_j; \hat S_{t_j}, \hat \Delta_j)$  permits to have an output in $[-1,1]^d$.
 \end{remark}

\begin{algorithm}[H]
\caption{\label{algoMeanVar1}Backward resolution for first local resolution algorithm (martingale case)}
\begin{algorithmic}[1]
\State $  \begin{array}{ll}
  U_N( \hat S_{t_{N}}(\omega), \hat \Delta_{N})  = g(S_T), \quad \forall  \hat \Delta_{N} \in [0,1]^d, 
\end{array} $
\For{$j = N-1,N-2, \dots, 1$}
\State   For $ x  \in U(0,1)^d $
\begin{eqnarray}
\label{eq:mainAlg1}
\theta_j^* =  \argmin_{ \theta } \Ebb \left[ \left(   U_{j+1}(  \hat S_{t_{j+1}}, \psi_{j+1}( \phi_j(\theta; \hat S_{t_j}, x))) - \phi_j(\theta; \hat S_{t_j}, x).(F_{t_{j+1}}-F_{t_j}) -    \hat V_j(\theta; \hat S_{t_j}, x) \right)^ 2  |  F_{t_j} \right]  ,
\end{eqnarray}
\hspace{1cm} where
\begin{eqnarray*}
 \phi_j(\theta; \hat S_{t_j}, x) = \left( \psi_j^{-1}(x) + l \tanh(\hat C_j(\theta,\hat S_{t_j}, x )) \right) 
\end{eqnarray*} 
\State $U_j( .,.)=  \hat V_j(\theta_j^*,.,.)$
\EndFor
\State At last:
\begin{eqnarray*}
\argmin_{ p \in \R, \Delta_0 \in [- l, l]} \Ebb \left[ (   U_{1}( \hat S_{t_{1}},  \psi_1(\Delta_0)) - C_0.(F_{t_{1}}-F_{t_0}) -    p )^ 2  \right]
\end{eqnarray*}
\end{algorithmic}
\end{algorithm}

\subsection{Second local algorithm}
The second algorithm can be seen as a path generalization of the first algorithm where at each time step an optimization is achieved to calculate the value function and the command at the current time step using the previously calculated commands. 
 In this algorithm  the gain functional $\bar R$ is updated $\omega$ by $\omega$. Then $\bar R$ satisfies at date $t_{i}$ with an asset value $S_{t_i}$  for an  investment $\Delta_{i-1}$ chosen at date $t_{i-1}$:
\begin{align*}
\bar R(t_i, S_{t_i}, \Delta_{i-1}) = &  g(S_T)  -  \sum_{k=1}^d \sum_{j=i}^{N-1}  \Delta_{j}^k  (F^k_{t_{j+1}}- F^k_{t_{j}}),\\
& = \bar R(t_{i+1}, S_{t_{i+1}}, \Delta_{i}) - \sum_{k=1}^d \Delta^k_{i}  (F^k_{t_{i+1}} - F^k_{t_i}) , 
\end{align*}
and, as shown in \cite{2017warin}, at the date $t_i$  the  optimal control $\Delta$  is  associated with the minimization  problem:
\begin{align*}
\min_{(V, \Delta) \in \R \times \R^d} \E\left[ (\bar R(t_{i+1}, S_{t_{i+1}}, \Delta) -   \sum_{k=1}^d  \Delta^k ( F^k_{t_{i+1}} -  F^k_{t_i}) - V )^ 2 | \Fc_{t_{i}}\right].
\end{align*}
This leads to the second algorithm \ref{algoMeanVar2}.
\begin{algorithm}[H]
\caption{\label{algoMeanVar2}Backward resolution for second local resolution algorithm}
\begin{algorithmic}[1]
\For{$j = N-1,N-2, \dots, 1$}
\State   For $ x  \in U(0,1)^d $
\begin{eqnarray}
\label{eq:mainAlg2}
\theta_j^* =  \argmin_{ \theta } \Ebb \left[ \left(   g(S_T)  -   \sum_{k=j}^{N-1}  \Delta_{k}.(F_{t_{k+1}}-F_{t_k}) -\hat V_j(\theta; \hat S_{t_j}, x) \right)^ 2  |  S_{t_j} \right]  ,
\end{eqnarray}
\hspace{1cm} where 
\begin{eqnarray*}
\Delta_{j} = & \phi_j(\theta; \hat S_{t_j}, x) \\
\Delta_{k+1}= & \phi_{k+1}( \theta^*_{k+1}, \hat S_{t_{k+1}}, \psi_{k+1}(\Delta_k)) \mbox{ for } k \in [j,N-2]
 \end{eqnarray*}
\hspace{1cm} and 
\begin{eqnarray*}
 \phi_k(\theta; \hat S_{t_k}, x) =  \psi_k^{-1}(x) + l \tanh(\hat C_k(\theta,\hat S_{t_k}, x ))  \mbox{ for } k \in [j,N-1]
\end{eqnarray*} 
\EndFor
\State At last:
\begin{eqnarray*}
\argmin_{ p \in \R, \Delta_{0} \in [- l, l]} \Ebb \left[ (    g(S_T)  -   \sum_{k=0}^{N-1}  \Delta_{k}.(F_{t_{k+1}}-F_{t_k}) -    p )^ 2  \right]
\end{eqnarray*}
\end{algorithmic}
\end{algorithm}
Each optimization is achieved using a stochastic gradient descent.
Notice that the second algorithm is far more costly than the first one as, at each time step, some command values have to be evaluated from the current time to the maturity of the asset to hedge.
\subsection{Parameters for the local algorithm}
We give the parameters used in the optimization process: 
\begin{itemize}
\item At each time step, a classical Feed Forward network of \textbf{four layers} (so one input layer, 2 hidden layers and one output layer) with \textbf{12 neurons} each is used. The three first layers use an ELU activation function while the output layer uses an identity activation function.
    \item The \textbf{batch size}, i.e. the number of simulations we use at each iteration to proceed an Adam gradient update is 2000. 
    \item At each time step the number of iterations used is limited to a number increasing with the dimension of the problem, from 5000 for the 4 dimensions problem to 25000 for problems which dimension strictly exceeds 4.
    \item The initial learning rate is $1e-3$.
\end{itemize}

\section{Numerical results in the transaction cost-free case and mean squared error}

In this section, we compare the three machine learning-based algorithms with a stochastic control based tool (\cite{stopt}) using a thin discretization to evaluate the optimal variance.

\subsection{Spread options payoff description}
\label{sec:spreadoption}
We use some spread option problem to compare the three algorithms.
The payoff in this section is defined for $d\geq 2$ by: 
\begin{eqnarray}
    g(S_T) & = & {\cal V}_T \left( F^1_T - \frac{1}{d-1} \sum_{i=2}^d F_T^i - K\right) ^+ . 
\end{eqnarray}
For all the cases we take the following parameters:
\begin{itemize}
    \item The maturity in days is equal to $T=90$ days,
    \item $K=10$,
    \item the number of hedging dates $N$ is taken equal to $14$ (but the control on last hedging date is trivial).
    \item $l^i$ the liquidity (i.e. the maximum quantity we can buy or sell) at each date is taken equal to $0.2$ for all underlying,
    \item $F^1_0=40$, $\sigma_{1, E}=0.004136$, $a_{1,E}=0.0002$ in days.
    \item the initial load associated to the option satisfies ${\cal V}_0=1$.
\end{itemize}
The three cases  take the following parameters:
\begin{enumerate}
    \item \underline{Case 1: $d=2, \sigma_{\cal V}=0$} \\
    This case is a four dimensional case (2 assets and 2 hedging positions) with:
    \begin{itemize}
        \item $F^2_0=30$, $\sigma_{2,E}=0.003137$, $a_{2,E}=0.0001$ in days.
        \item $\rho_{1,2}=0.7$ is the correlation between the two assets.
    \end{itemize}
    \item \underline{Case 2: $d=2$}\\
    This is a 5 dimensional case, with the same parameters as in the first case but with a varying load with parameters
    $\sigma_{\cal V}=0.02$, $a_{\cal V}=0.02$ in days. The correlation between each of the tradeable assets and the non-tradable asset  $\cal V$ is equal to $0.2$. Note that this is the only case where $\sigma_v>0$.
    \item \underline{Case 3: $d=3$, $\sigma_{\cal V}=0$}\\
    This is a case in dimension $6$ with
    \begin{itemize}
        \item $F^2_0=35$, $\sigma_{2,E}=0.003137$, $a_{2,E}=0.0001$ in days.
        \item $F^3_0=25$, $\sigma_{3,E}=0.005136$, $a_{3,E}=0.0001$ in days.
        \item The correlation between asset $i$ and $j$ is noted $\rho_{i,j}$ and satisfies: $\rho_{1,2}=0.7, \rho_{1,3}= 0.3,  \rho_{2,3}=0.5$.
    \end{itemize}
\end{enumerate}

\subsubsection{Numerical results}
\label{sec:resultsmeansquare}
In Table \ref{tab:marketspread}, the variance obtained on 100 000 common simulations are given for the 3 algorithms and compared to the variance obtained by the StOpt library. Notice that due to the size of the problem the case 3 is not totally converged with the StOpt library.\\
For local algorithm 1 and 2, we run the optimization 10 times and take the best variance obtained. \\
The global algorithm is far more effective in term of computing time than the local algorithm as 10 000 iterations runs in 220 s on the graphic card of a core I3 laptop while algorithm 2 and 3  can take some hours for the case 3.

\begin{figure}[h!]
\begin{minipage}[t]{0.49\linewidth}
  \centering
 \includegraphics[width=\textwidth]{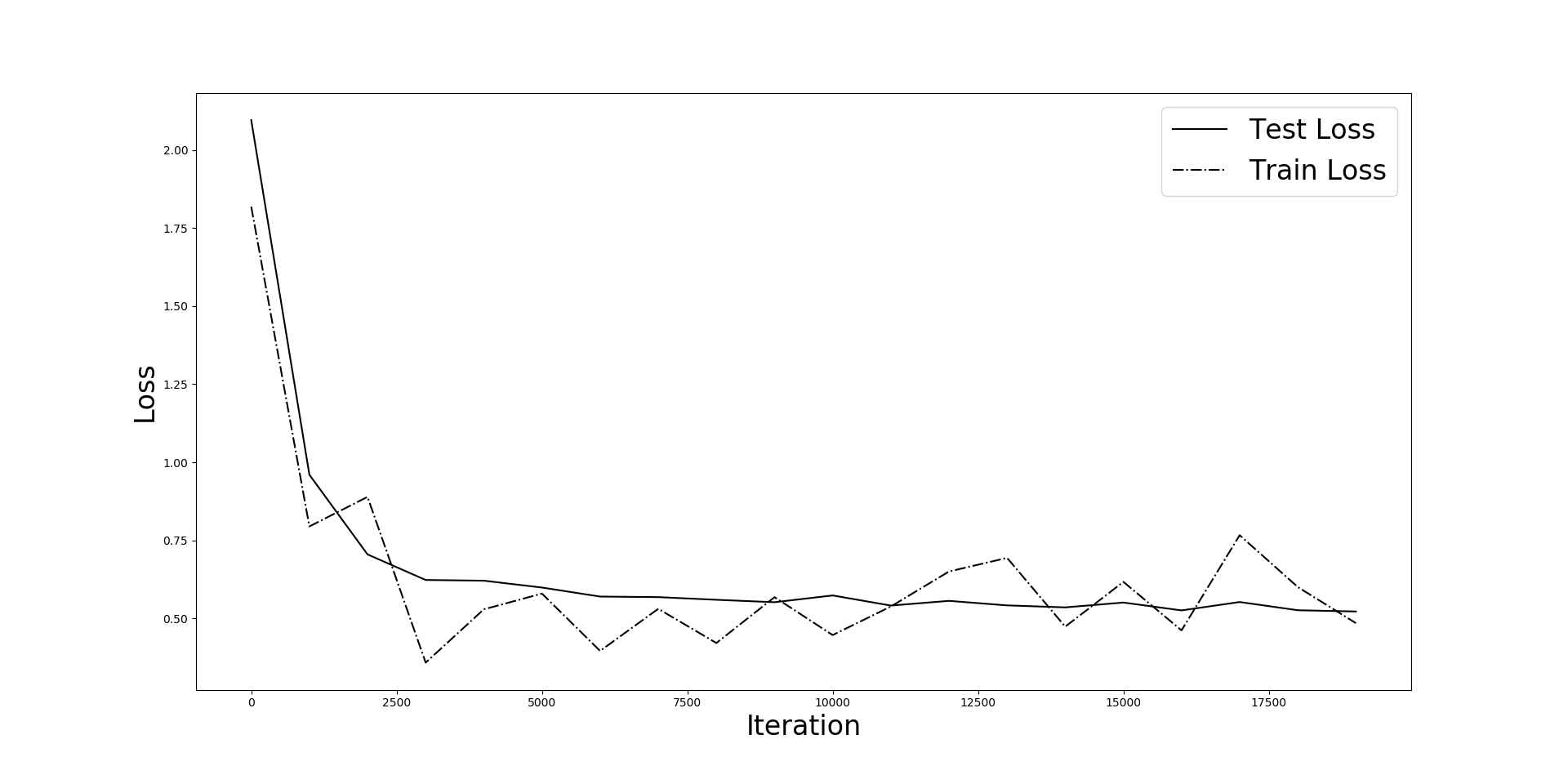}
 \caption*{Case 1. }
 \end{minipage}
\begin{minipage}[t]{0.49\linewidth}
  \centering
 \includegraphics[width=\textwidth]{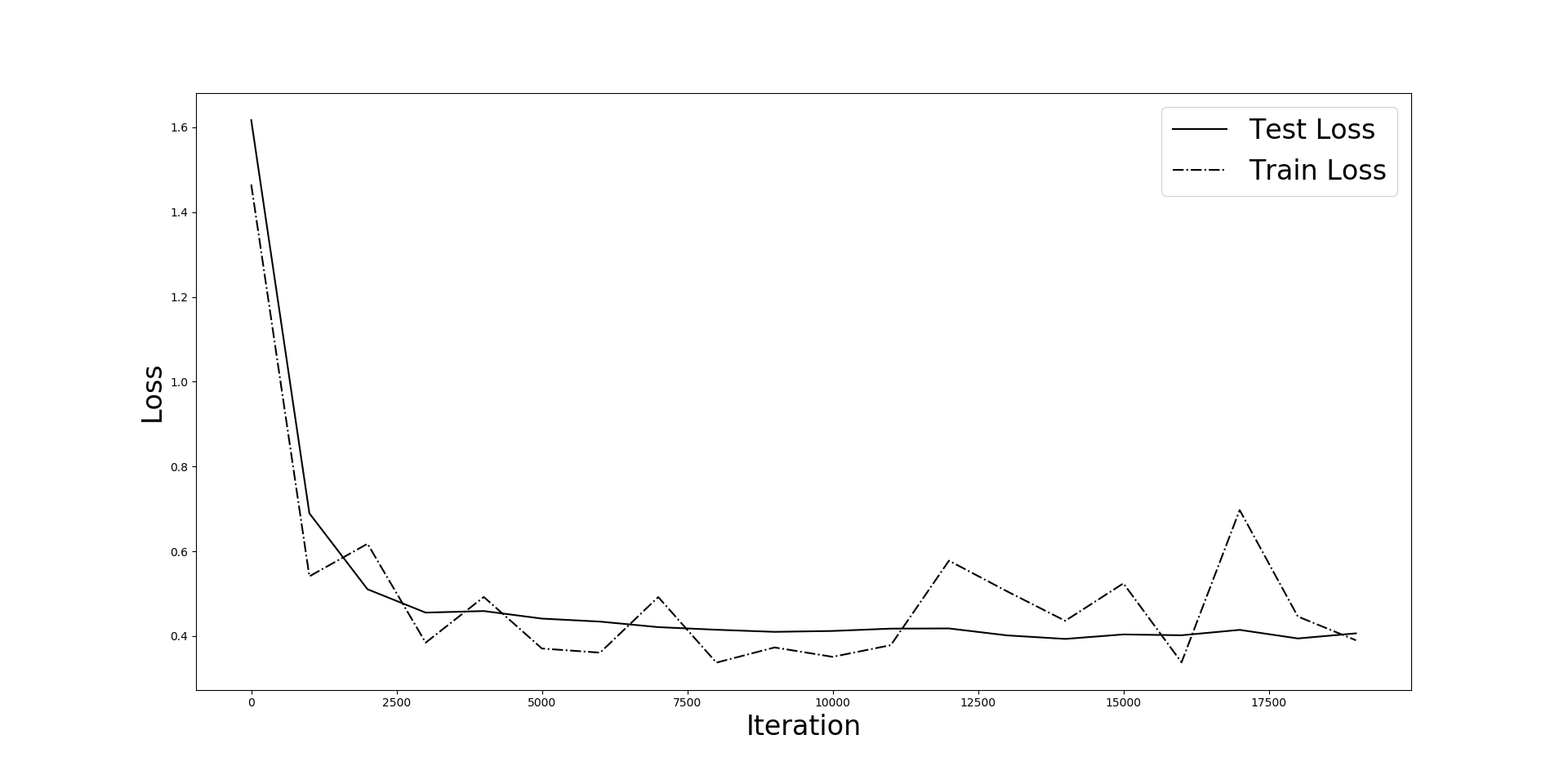}
 \caption*{Case 2.}
 \end{minipage}
 \begin{minipage}[t]{0.49\textwidth}
  \centering
 \includegraphics[width=\textwidth]{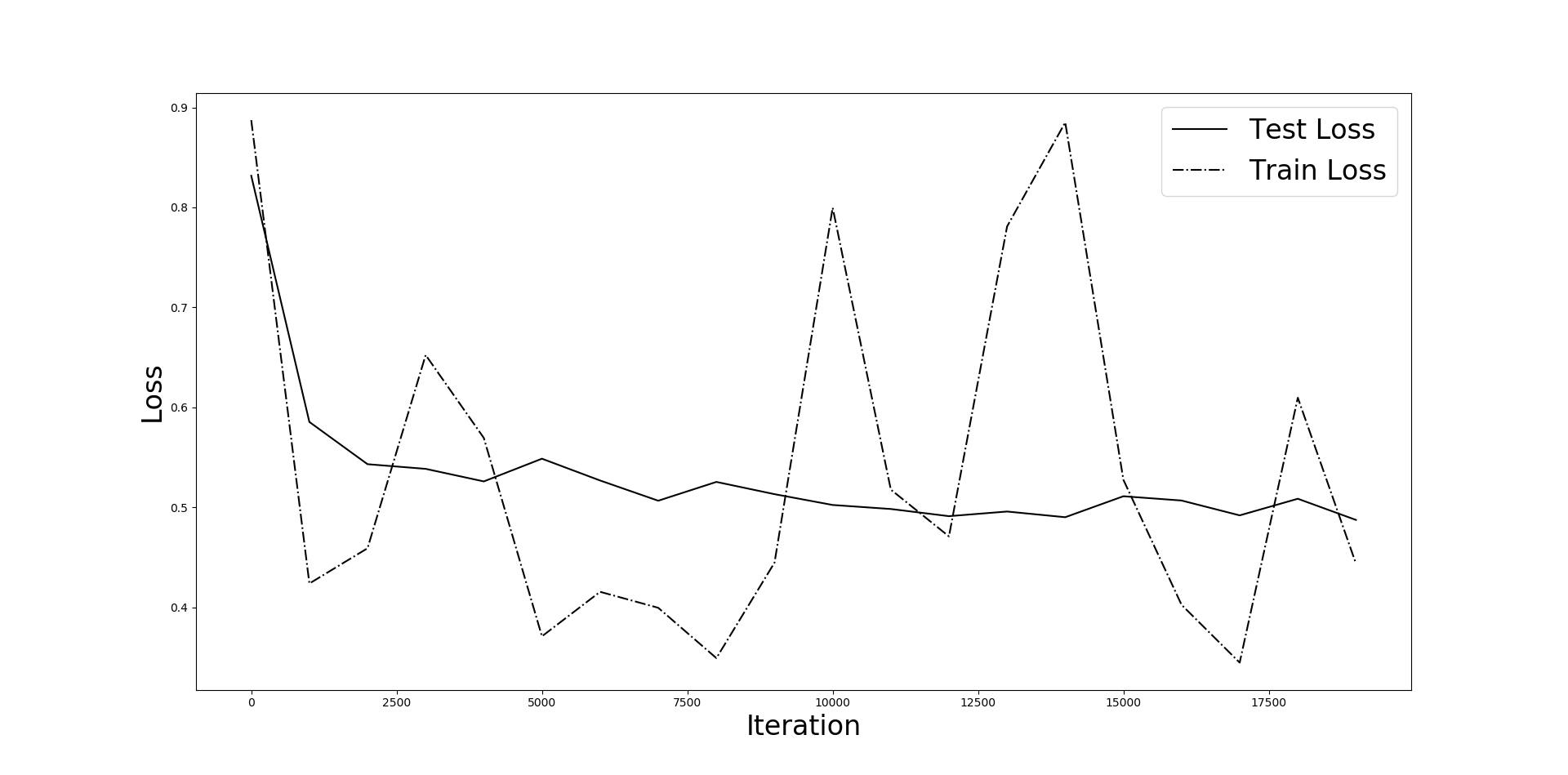}
 \caption*{Case 3.}
 \end{minipage}
  \begin{minipage}[t]{0.49\textwidth}
  \centering
 \includegraphics[width=\textwidth]{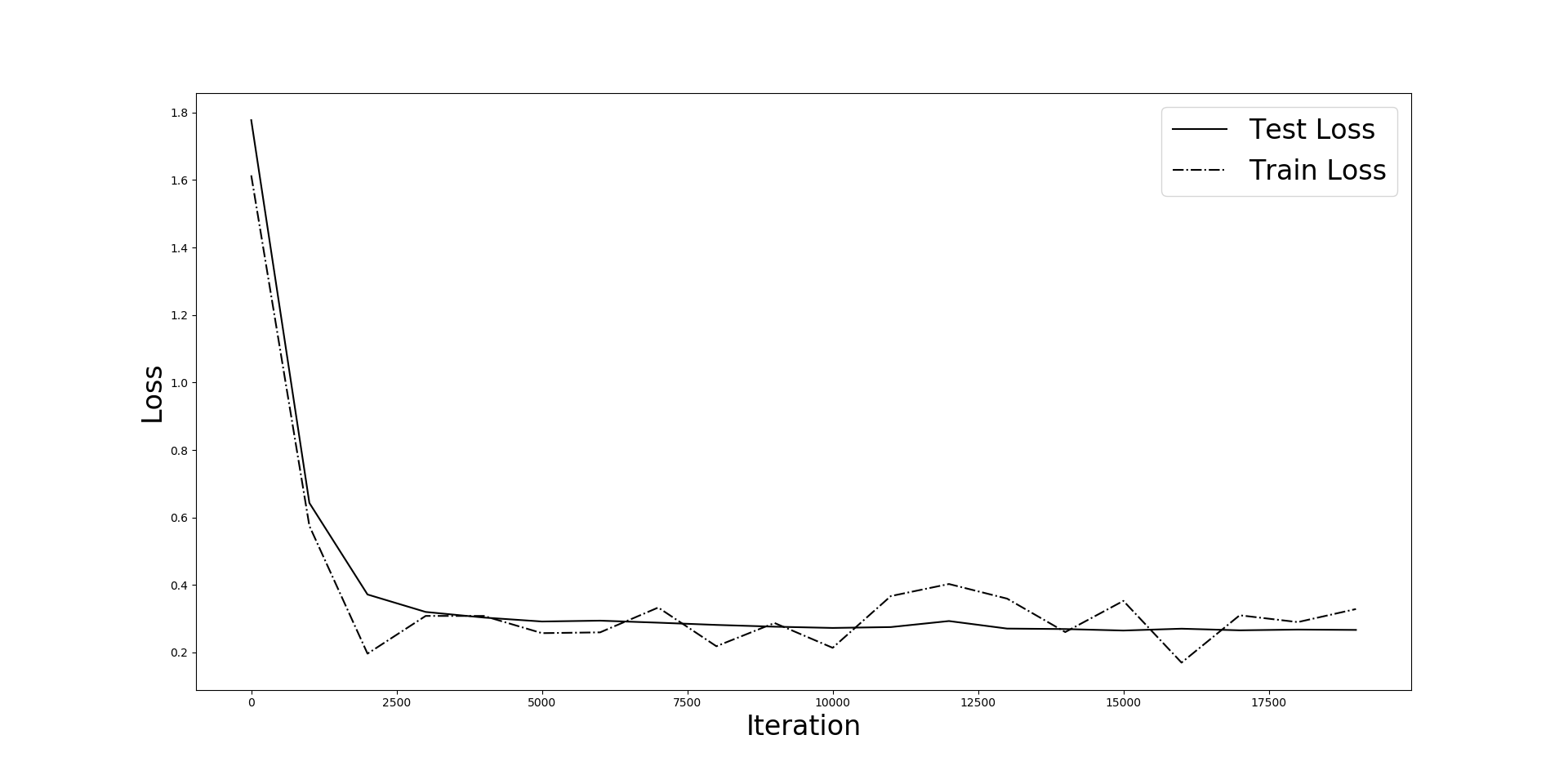}
 \caption*{Case 2. with two time more time steps}
 \end{minipage}

\caption{Loss functions for the Global NN algorithm.}
    \label{fig:lossmarketspread}
 \end{figure}

\begin{table}[!ht]
\centering
\begin{tabular}{|l|l|l|l|}
\hline
\textbf{Mean Squared Error}  & \textbf{Case 1}  & \textbf{Case 2} &\textbf{ Case 3}   \\ \hline 
Unhedged Portfolio & 8.3058 & 8.5250 & 10.5960 \\ \hline \hline 
Hedged with StOpt & 0.3931 &  0.5160 & 0.4983 \\ \hline
Hedged with Global Algo  &  0.3920  & 0.5205  & 0.4852  \\ \hline
Hedged with Algo 1 & 0.3971&   0.5168 & 0.4763 \\ \hline 
Hedged with Algo 2 & 0.3912& 0.5183 & 0.4943\\ \hline 
\end{tabular}
\caption{Mean Square comparison between, NN-based algorithms and stochastic control algorithm}
\label{tab:marketspread}
\end{table}
In Figure \ref{fig:lossmarketspread}, the losses for the market spread and for the Global NN algorithm are plotted.

\begin{figure}[h!]
 \begin{minipage}[b]{0.32\linewidth}
  \centering
 \includegraphics[width=\textwidth]{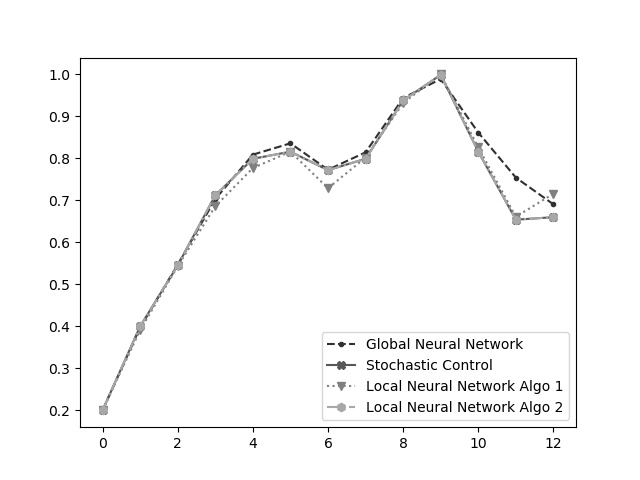}
 \caption*{Future 1 Sim Nb 1}
 \end{minipage}
  \begin{minipage}[b]{0.32\linewidth}
  \centering
 \includegraphics[width=\textwidth]{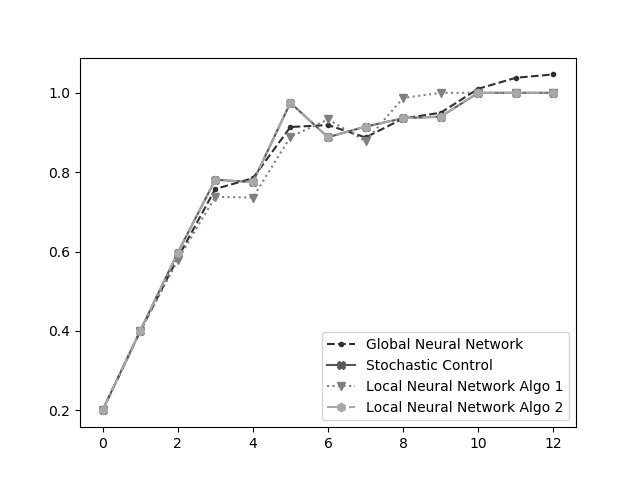}
 \caption*{Future 1 Sim Nb 2}
 \end{minipage}
   \begin{minipage}[b]{0.32\linewidth}
  \centering
 \includegraphics[width=\textwidth]{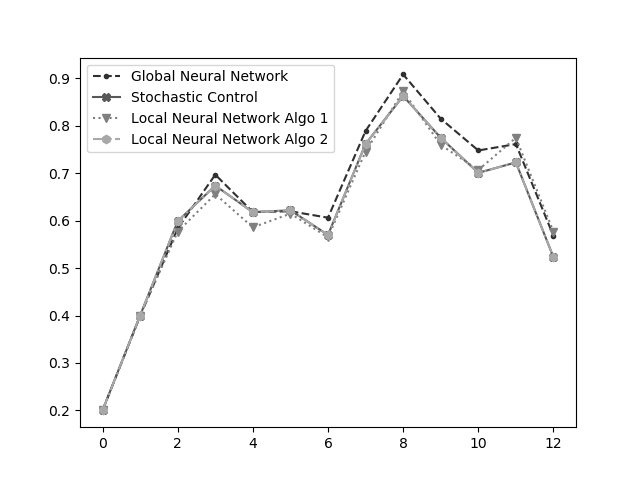}
 \caption*{Future 1 Sim Nb 3}
 \end{minipage}
  \begin{minipage}[b]{0.32\linewidth}
  \centering
 \includegraphics[width=\textwidth]{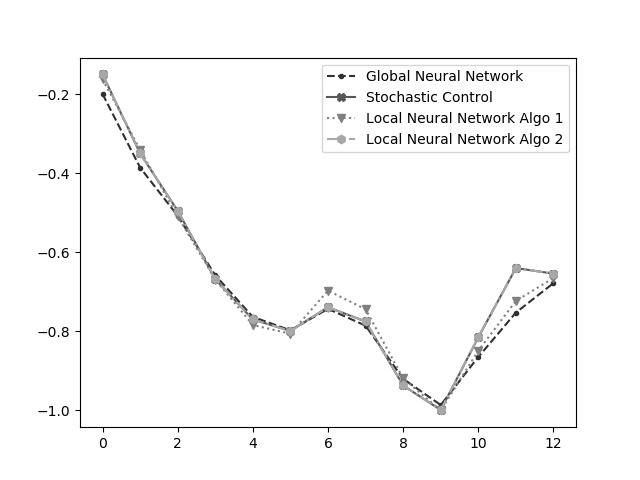}
 \caption*{Future 2 Sim Nb 1}
 \end{minipage}
  \begin{minipage}[b]{0.32\linewidth}
  \centering
 \includegraphics[width=\textwidth]{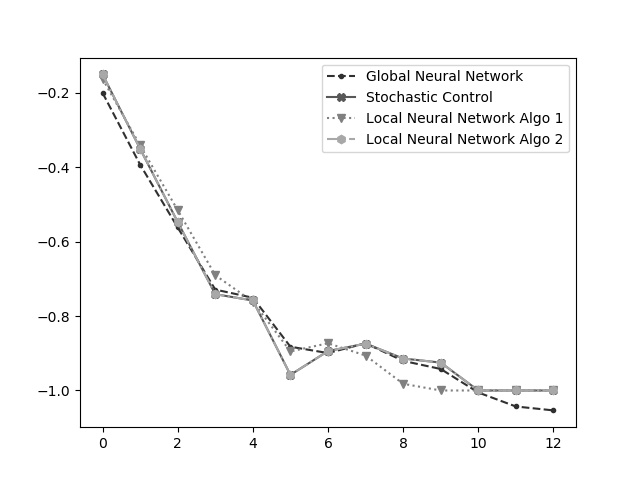}
 \caption*{Future 2 Sim Nb 2}
 \end{minipage}
   \begin{minipage}[b]{0.32\linewidth}
  \centering
 \includegraphics[width=\textwidth]{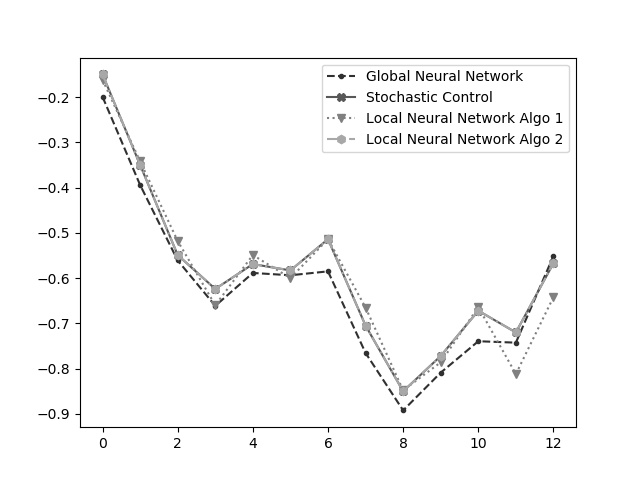}
 \caption*{Future 2 Sim Nb 3}
 \end{minipage}
 \caption{Delta for Case 1.}
 \label{fig:deltaCase1}
 \end{figure}

\begin{figure}[h!]
 \begin{minipage}[b]{0.32\linewidth}
  \centering
 \includegraphics[width=\textwidth]{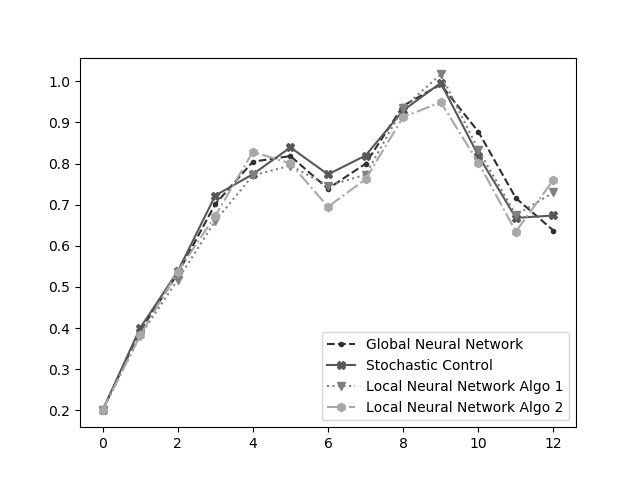}
 \caption*{Future 1 Sim Nb 1}
 \end{minipage}
  \begin{minipage}[b]{0.32\linewidth}
  \centering
 \includegraphics[width=\textwidth]{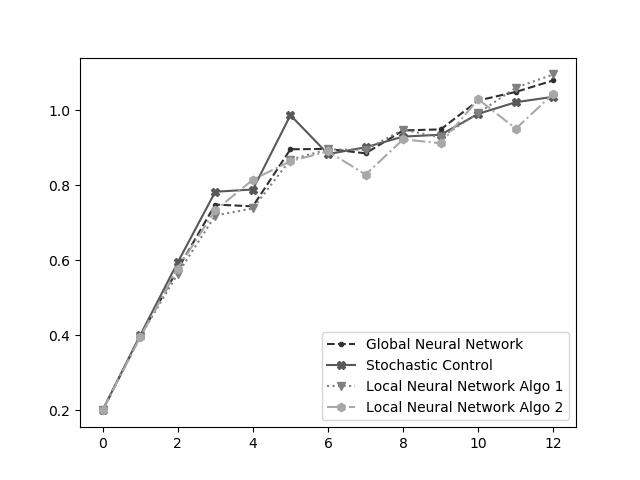}
 \caption*{Future 1 Sim Nb 2}
 \end{minipage}
   \begin{minipage}[b]{0.32\linewidth}
  \centering
 \includegraphics[width=\textwidth]{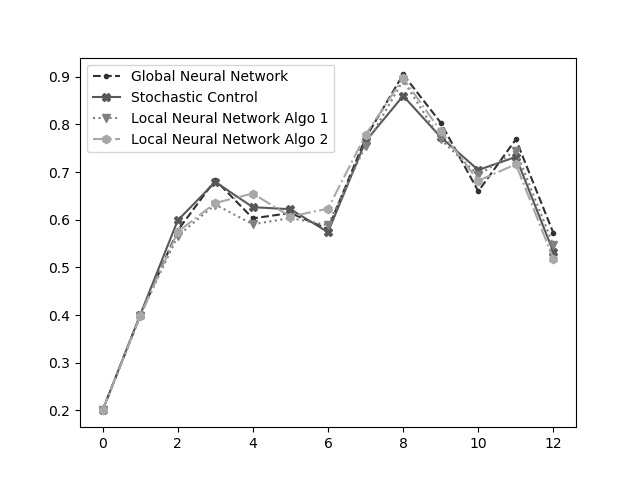}
 \caption*{Future 1 Sim Nb 3}
 \end{minipage}
  \begin{minipage}[b]{0.32\linewidth}
  \centering
 \includegraphics[width=\textwidth]{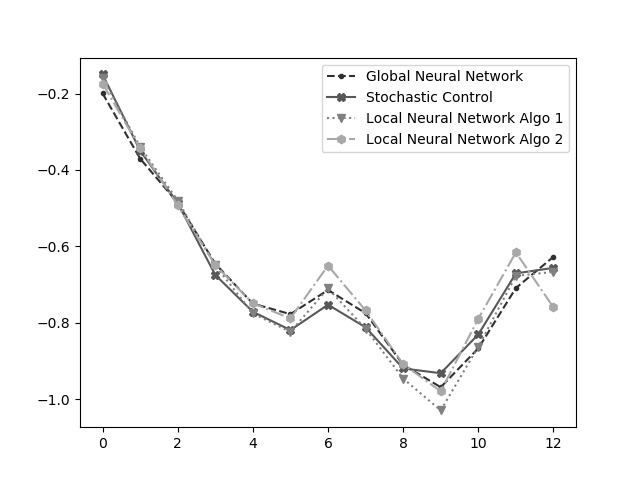}
 \caption*{Future 2 Sim Nb 1}
 \end{minipage}
  \begin{minipage}[b]{0.32\linewidth}
  \centering
 \includegraphics[width=\textwidth]{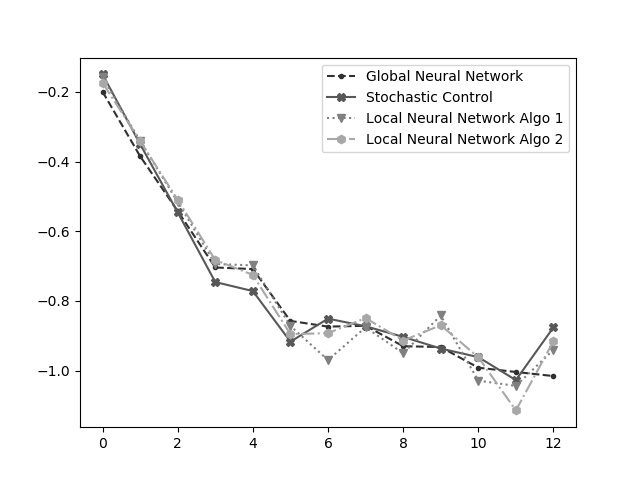}
 \caption*{Future 2 Sim Nb 2}
 \end{minipage}
   \begin{minipage}[b]{0.32\linewidth}
  \centering
 \includegraphics[width=\textwidth]{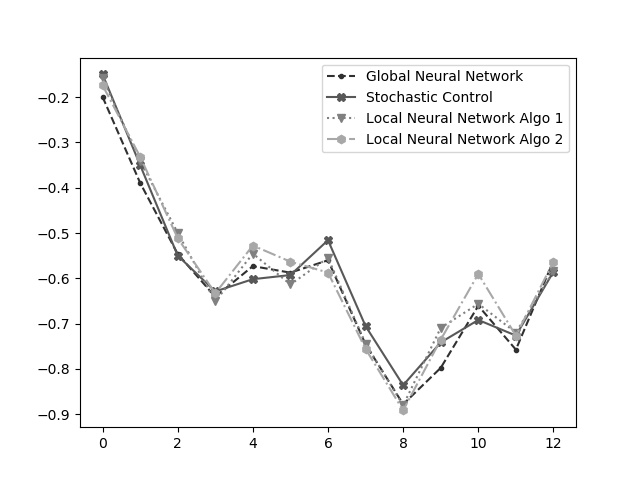}
 \caption*{Future 2 Sim Nb 3}
 \end{minipage}
 \caption{Delta for Case 2.}
 \label{fig:deltaCase2}
 \end{figure}

\begin{figure}[h!]
 \begin{minipage}[b]{0.32\linewidth}
  \centering
 \includegraphics[width=\textwidth]{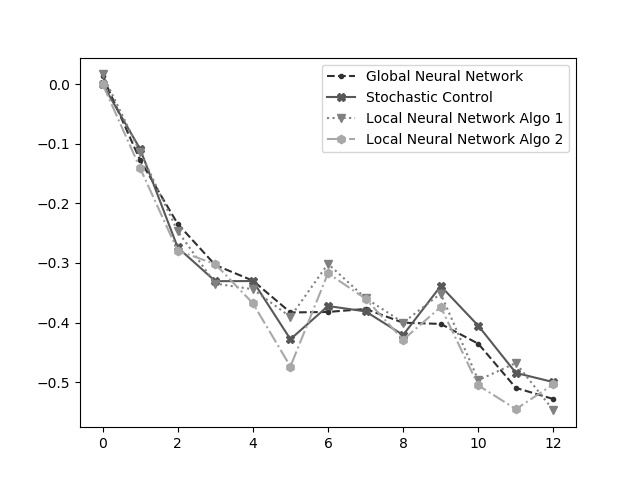}
 \caption*{Future 2 Sim Nb 1}
 \end{minipage}
  \begin{minipage}[b]{0.32\linewidth}
  \centering
 \includegraphics[width=\textwidth]{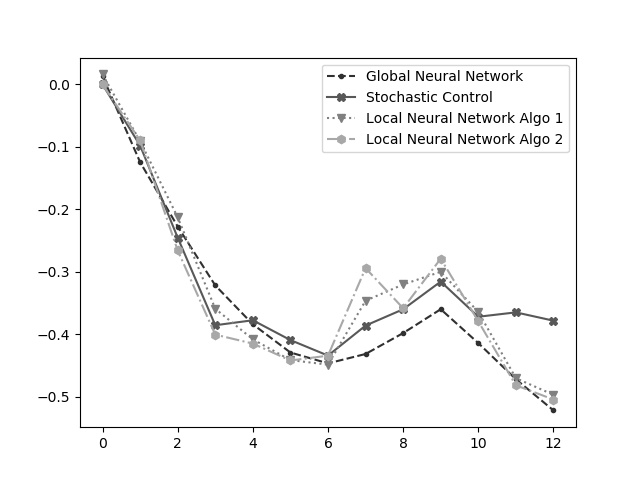}
 \caption*{Future 2 Sim Nb 2}
 \end{minipage}
   \begin{minipage}[b]{0.32\linewidth}
  \centering
 \includegraphics[width=\textwidth]{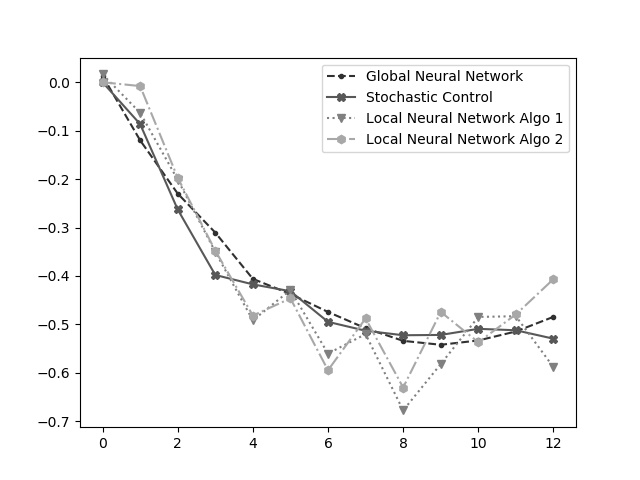}
 \caption*{Future 2 Sim Nb 3}
 \end{minipage}
  \begin{minipage}[b]{0.32\linewidth}
  \centering
 \includegraphics[width=\textwidth]{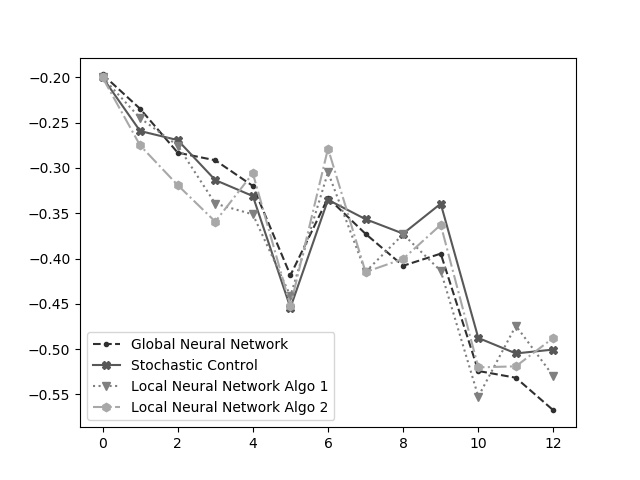}
 \caption*{Future 3 Sim Nb 1}
 \end{minipage}
  \begin{minipage}[b]{0.32\linewidth}
  \centering
 \includegraphics[width=\textwidth]{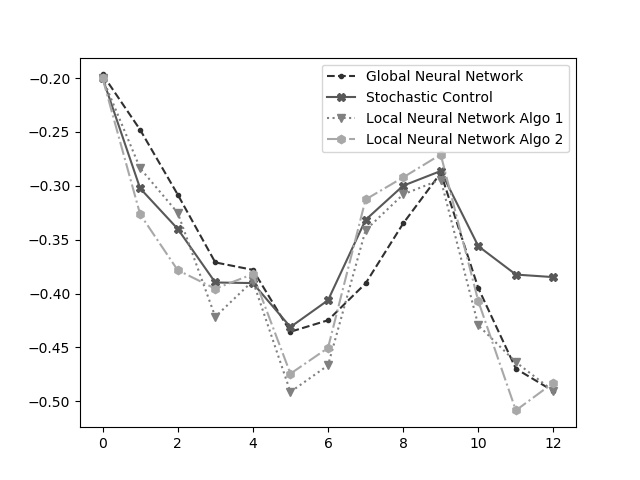}
 \caption*{Future 3 Sim Nb 2}
 \end{minipage}
   \begin{minipage}[b]{0.32\linewidth}
  \centering
 \includegraphics[width=\textwidth]{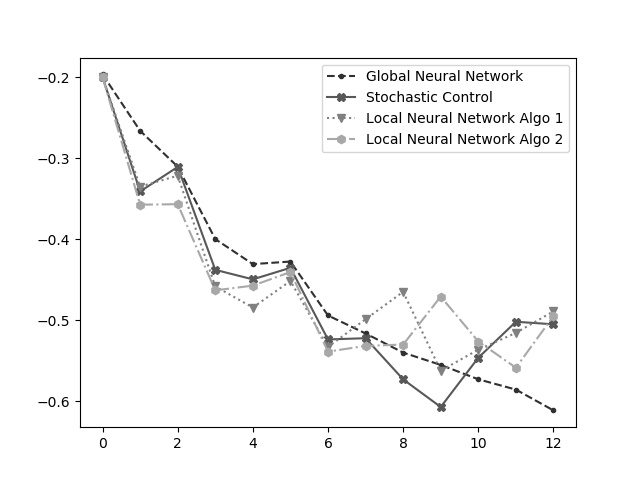}
 \caption*{Future 3 Sim Nb 3}
 \end{minipage}
 \caption{Delta for Case 3.}
 \label{fig:deltaCase3}
 \end{figure}

\FloatBarrier
As shown in Figures \ref{fig:deltaCase1}, \ref{fig:deltaCase2} and \ref{fig:deltaCase3} the Deltas for the 2 and 3 markets spread follow the same shape for the four algorithms. 

 The numerical results indicate that the global algorithm and local algorithm give similar results. We observe that, using 10 runs, the local algorithms gives similar results in the low dimension, but as the dimension increases, the results obtained may differ a lot meaning that the optimizer is often trapped in a local minimum solution far from the result. Besides the number of iterations to use at each step has to be increased a lot with the dimension leading to a non-competitive running time compared to the global algorithm.
\\
The global algorithm is still very effective in dimension 6 and, being able to solve the problem very quickly, is a candidate to give a method solving problems in very high dimensions. \\ One question that arises is how the three neural-network-based algorithms perform when the number of decisions i.e. the number of hedging dates increases. To increase the number of hedging dates we can increase the maturity $T$ while keeping the same distance between two hedging dates. Due to the mean reverting nature of the chosen models a more complex case consists in keeping $T=90$ days while increasing the number of hedging dates. In Table \ref{tab:marketspreadmoredates} we compute the error of the four algorithms with 28 (instead of 14 previously) hedging dates and a liquidity of 0.15 (instead of 0.20) units per date. The three approaches are effective in term of accuracy.
The time spent with the local algorithm 2 explodes due to the resimulation at each date of the optimal strategy until maturity.
\begin{table}[!ht]
\centering
\begin{tabular}{|l|l|l|l|l|}
\hline
\textbf{ Stochastic Control  }& \textbf{Global }& \textbf{Algo 1 }& \textbf{Algo 2 }\\ \hline 
 0.271 & 0.265 & 0.259 & 0.262  \\ \hline
 
\end{tabular}
\caption{Mean Squared error on Case 1 with 24 hedging dates and a liquidity of 0.15.}
\label{tab:marketspreadmoredates}
\end{table}

\subsection{Testing different risk criteria} 
\label{sec:riskcriteria}
\begin{figure}[h!]
\begin{minipage}[b]{1.\textwidth}

 \includegraphics[width=\textwidth]{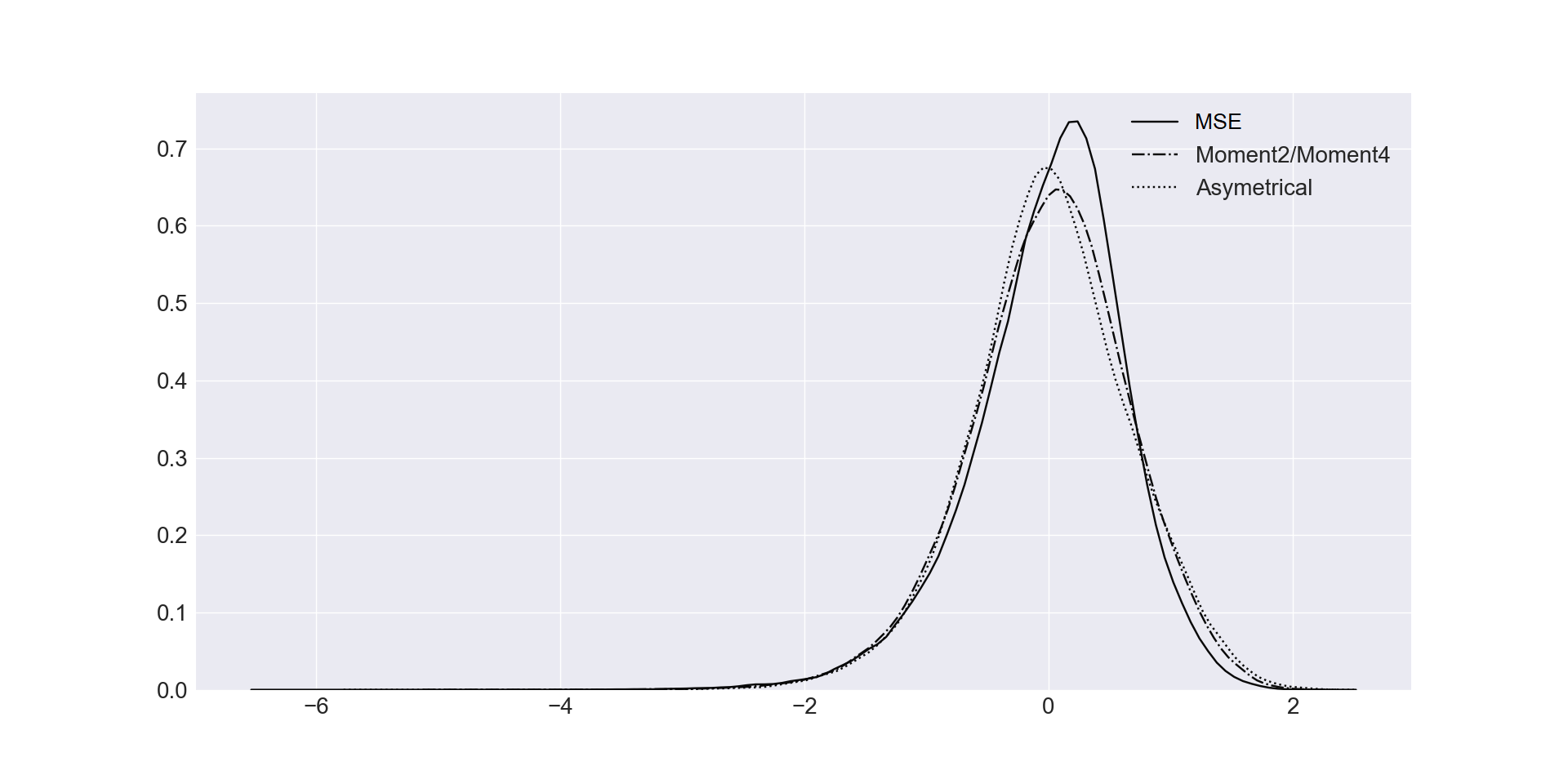}
 \caption*{}
 \end{minipage}
  \begin{minipage}[b]{0.49\linewidth}
  \centering
 \includegraphics[width=\textwidth]{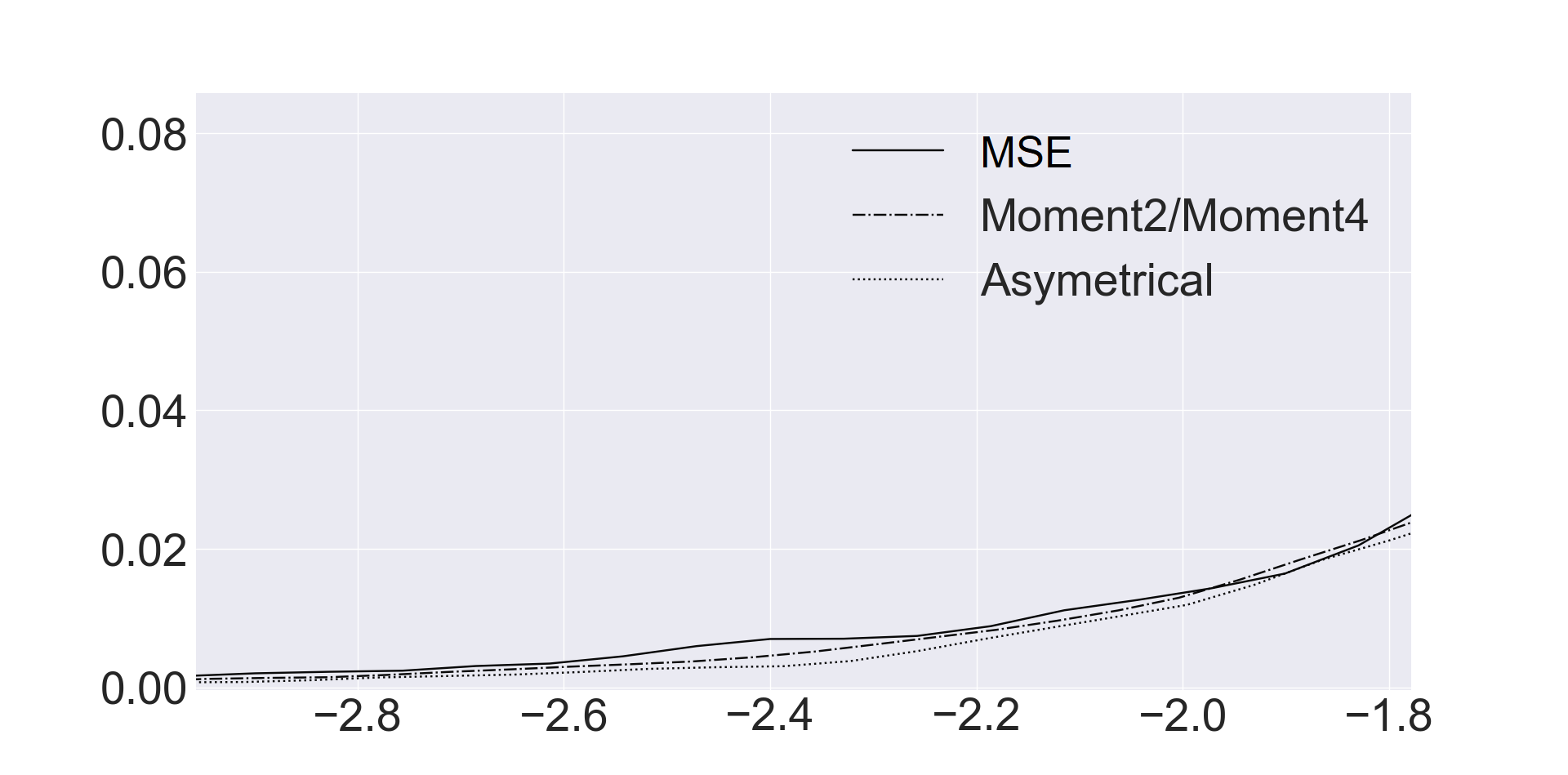}
 \caption*{Zoom on left hand tail}
 \end{minipage}
  \begin{minipage}[b]{0.49\linewidth}
  \centering
 \includegraphics[width=\textwidth]{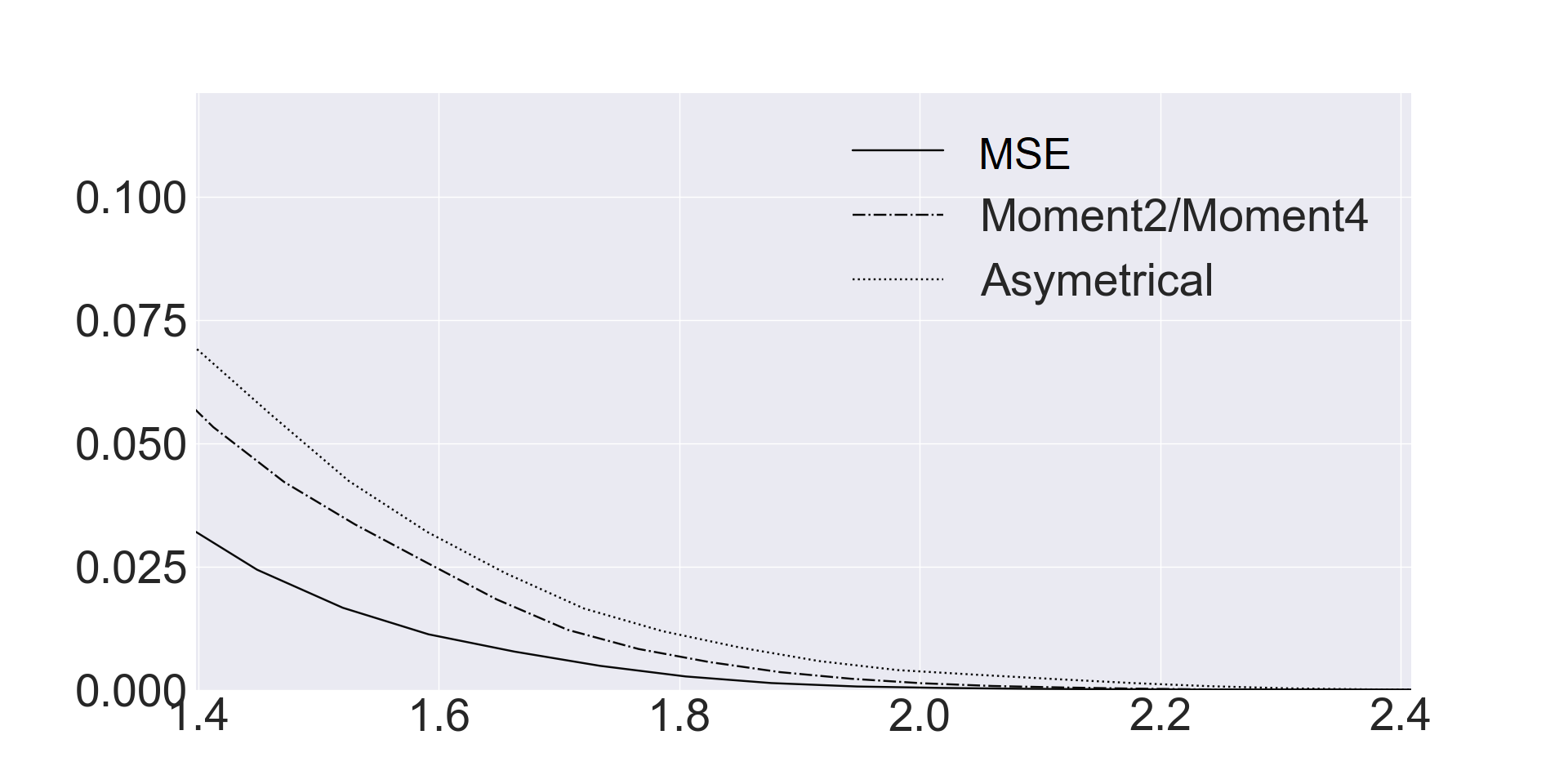}
 \caption*{Zoom on right hand tail}
 \end{minipage}
 \caption{Distribution of the hedged portfolio for Case 1 and different risk criterion - Zoom on the tails}
  \label{fig:riskcase1}
 \end{figure}

\begin{figure}[h!]
\begin{minipage}[b]{1.\textwidth}
  \centering
 \includegraphics[width=\textwidth]{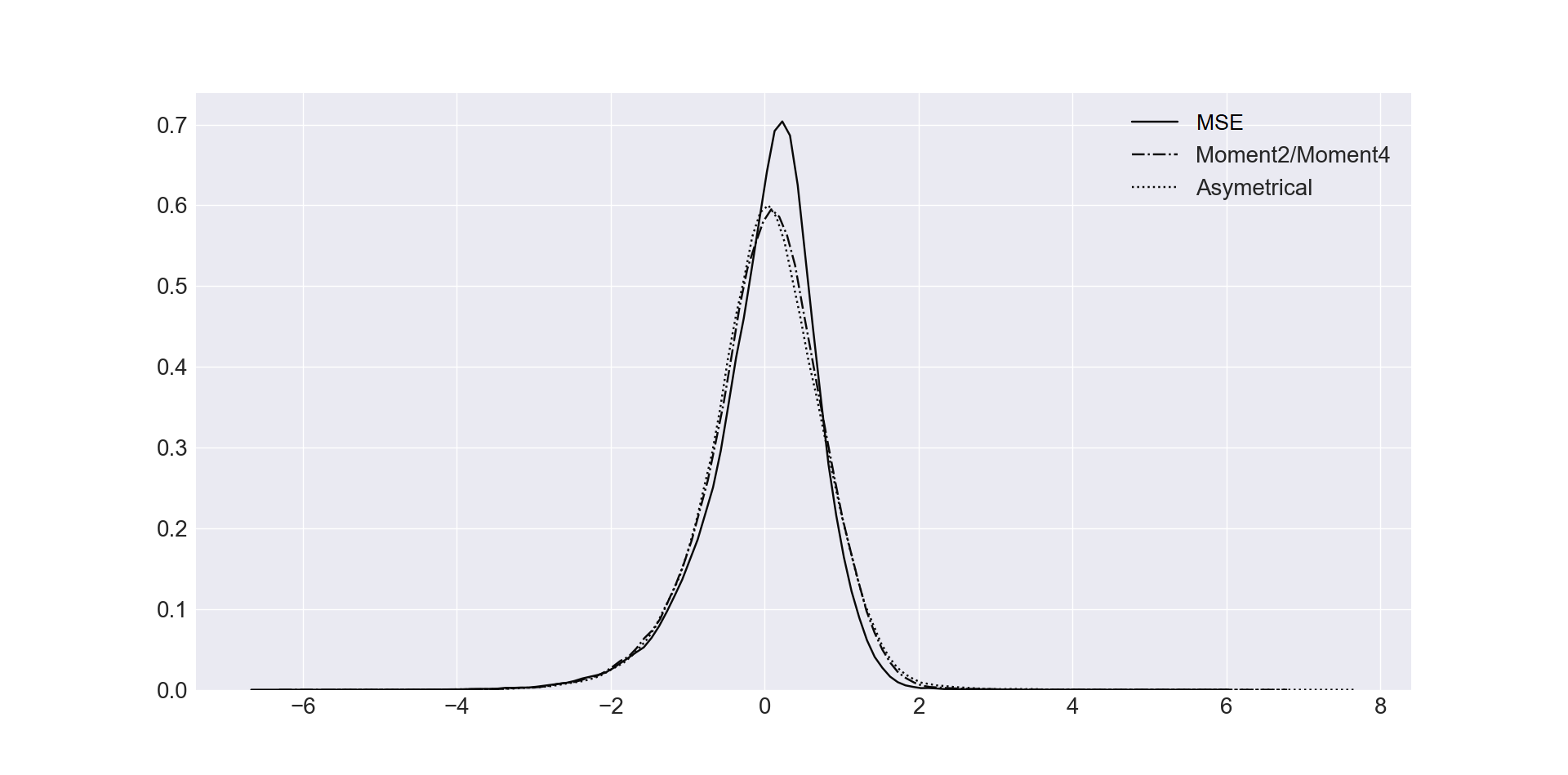}
 \caption*{}
 \end{minipage}
  \begin{minipage}[b]{0.49\linewidth}
  \centering
 \includegraphics[width=\textwidth]{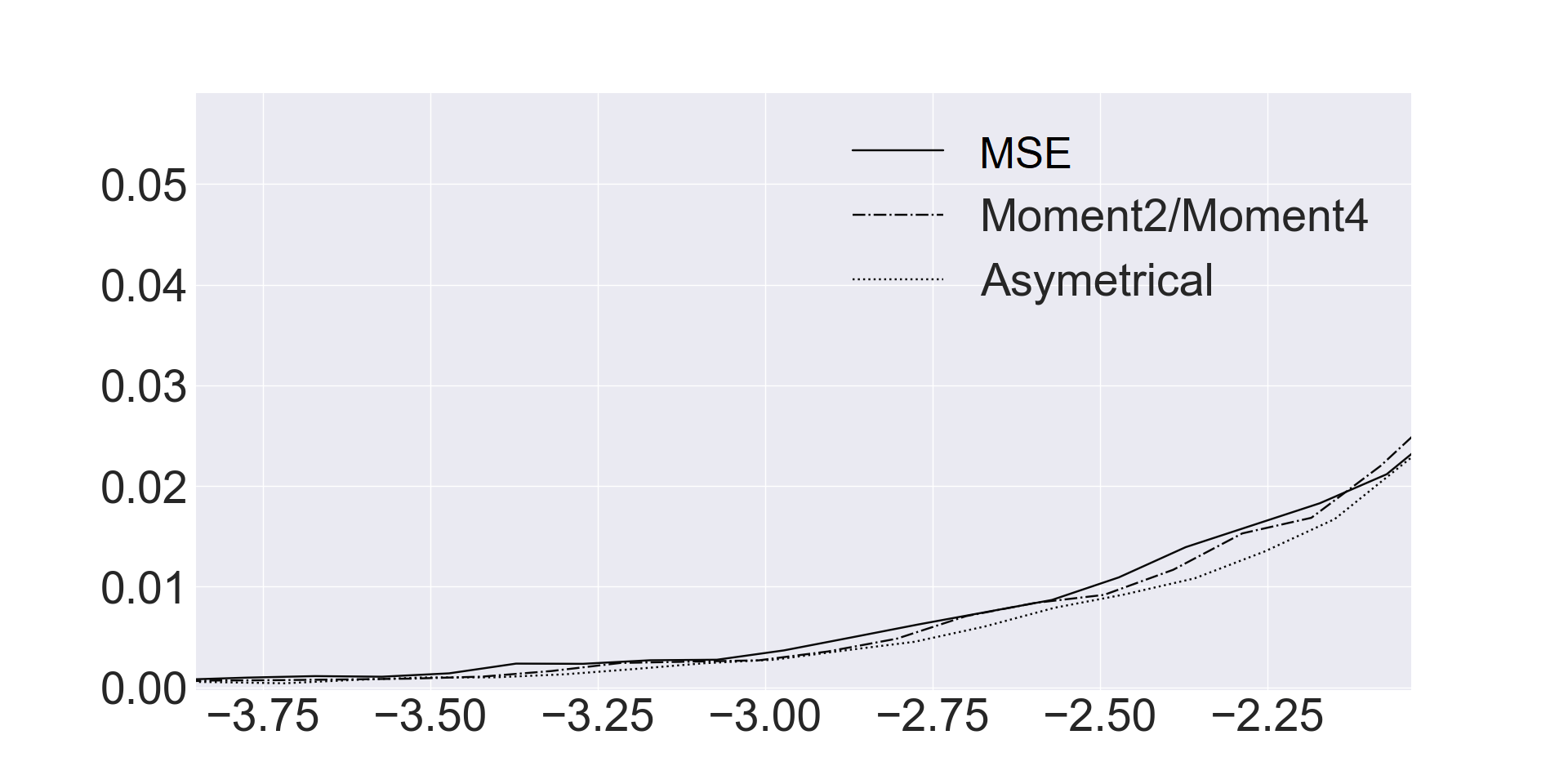}
 \caption*{Zoom on left hand tail}
 \end{minipage}
  \begin{minipage}[b]{0.49\linewidth}
  \centering
 \includegraphics[width=\textwidth]{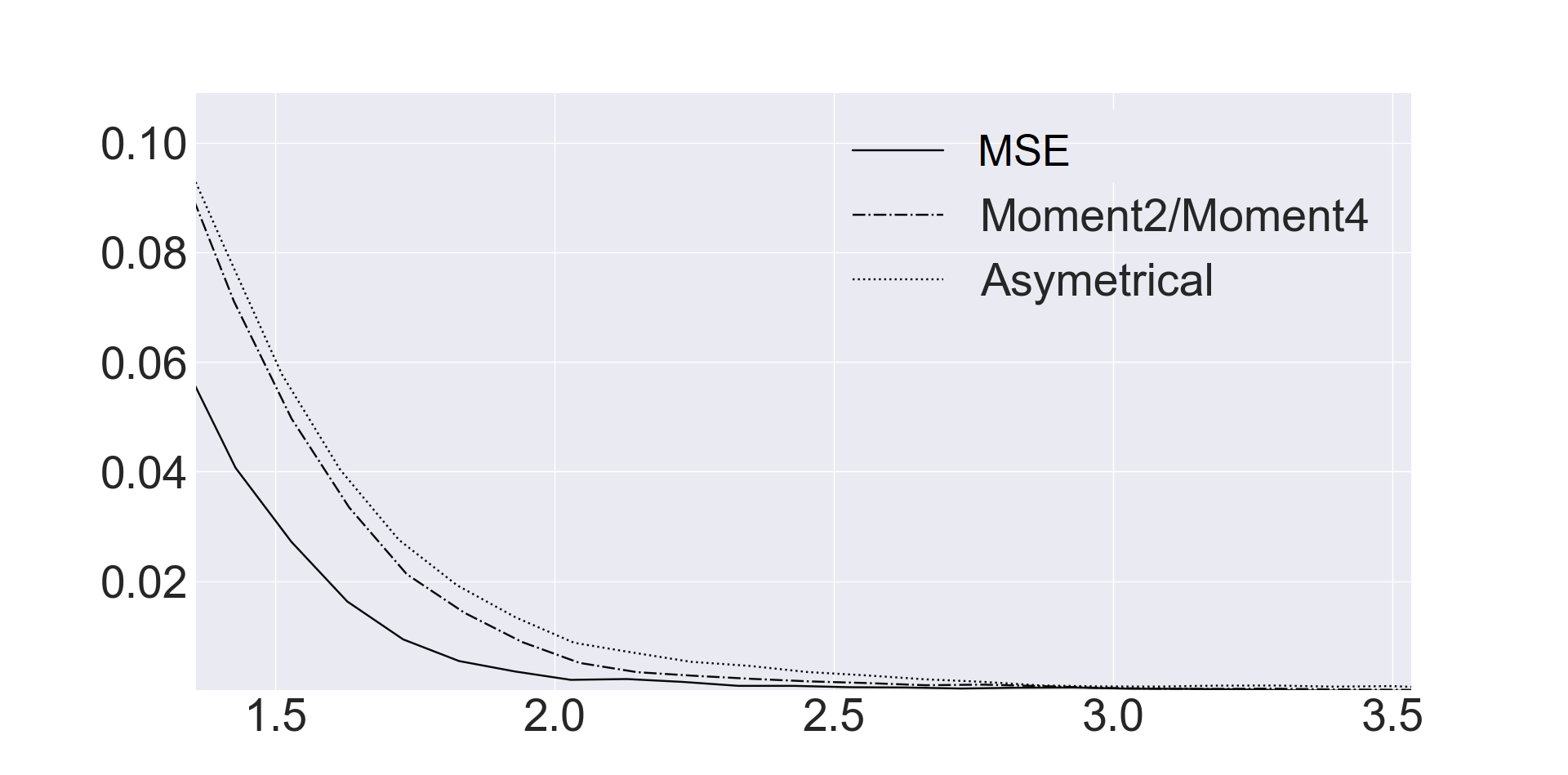}
 \caption*{Zoom on right hand tail}
 \end{minipage}

 \caption{Distribution of the hedged portfolio for Case 2 and different risk criterion - Zoom on the tails}
  \label{fig:riskcase2}
 \end{figure}

\begin{figure}[h!]
\begin{minipage}[b]{1.\textwidth}
  \centering
 \includegraphics[width=\textwidth]{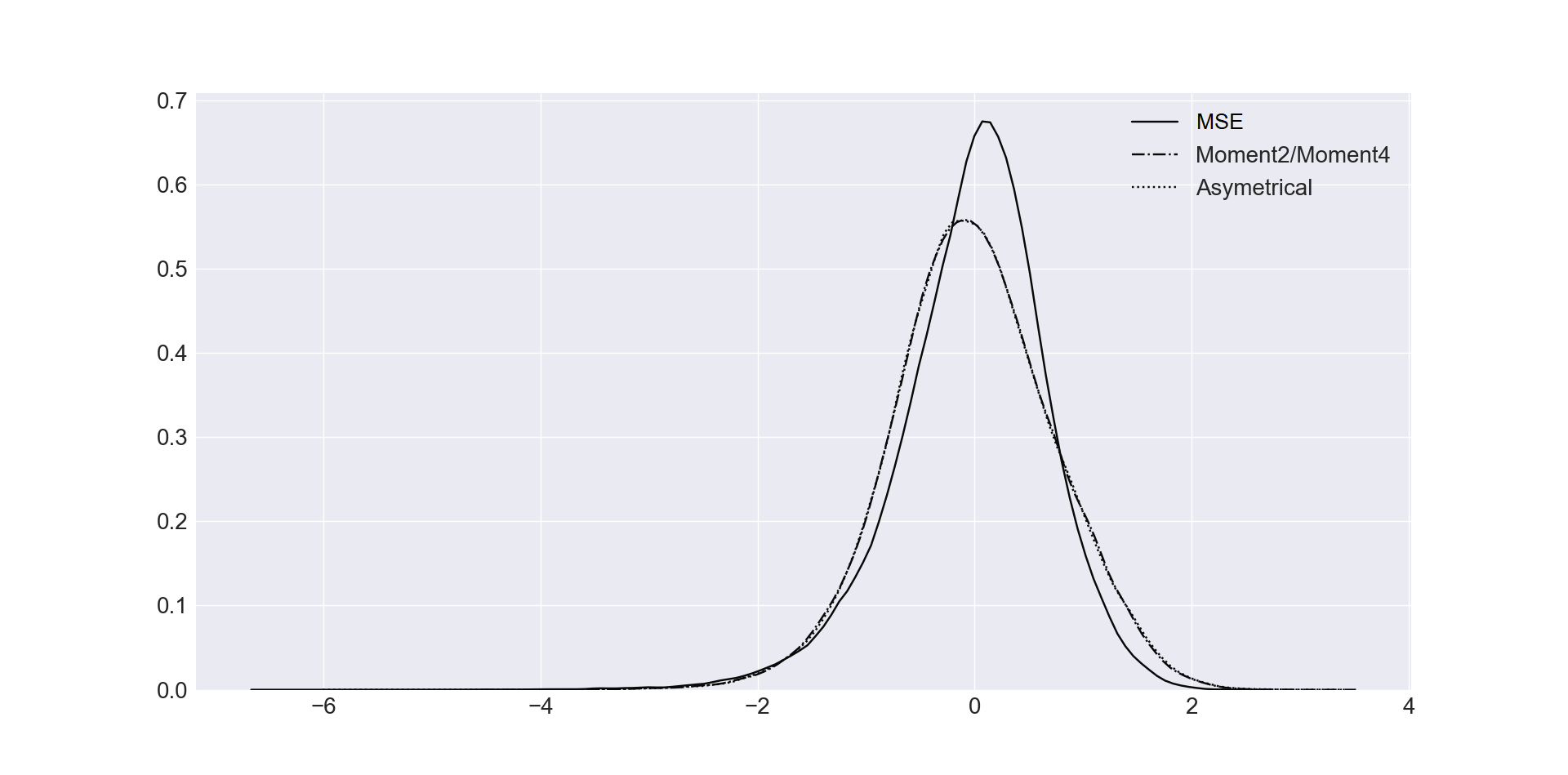}
 \caption*{}
 \end{minipage}
  \begin{minipage}[b]{0.49\linewidth}
  \centering
 \includegraphics[width=\textwidth]{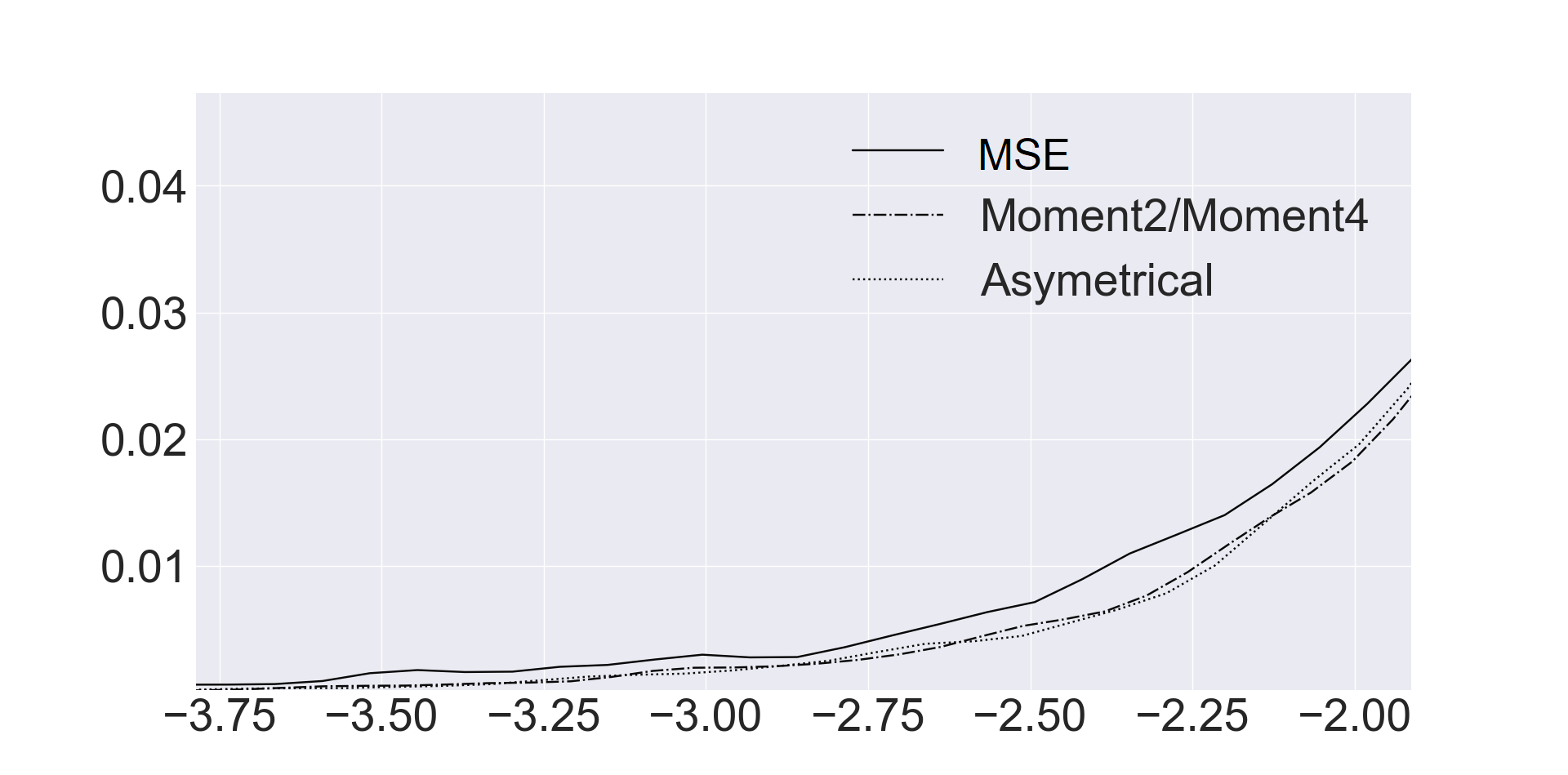}
 \caption*{Zoom on left hand tail}
 \end{minipage}
  \begin{minipage}[b]{0.49\linewidth}
  \centering
 \includegraphics[width=\textwidth]{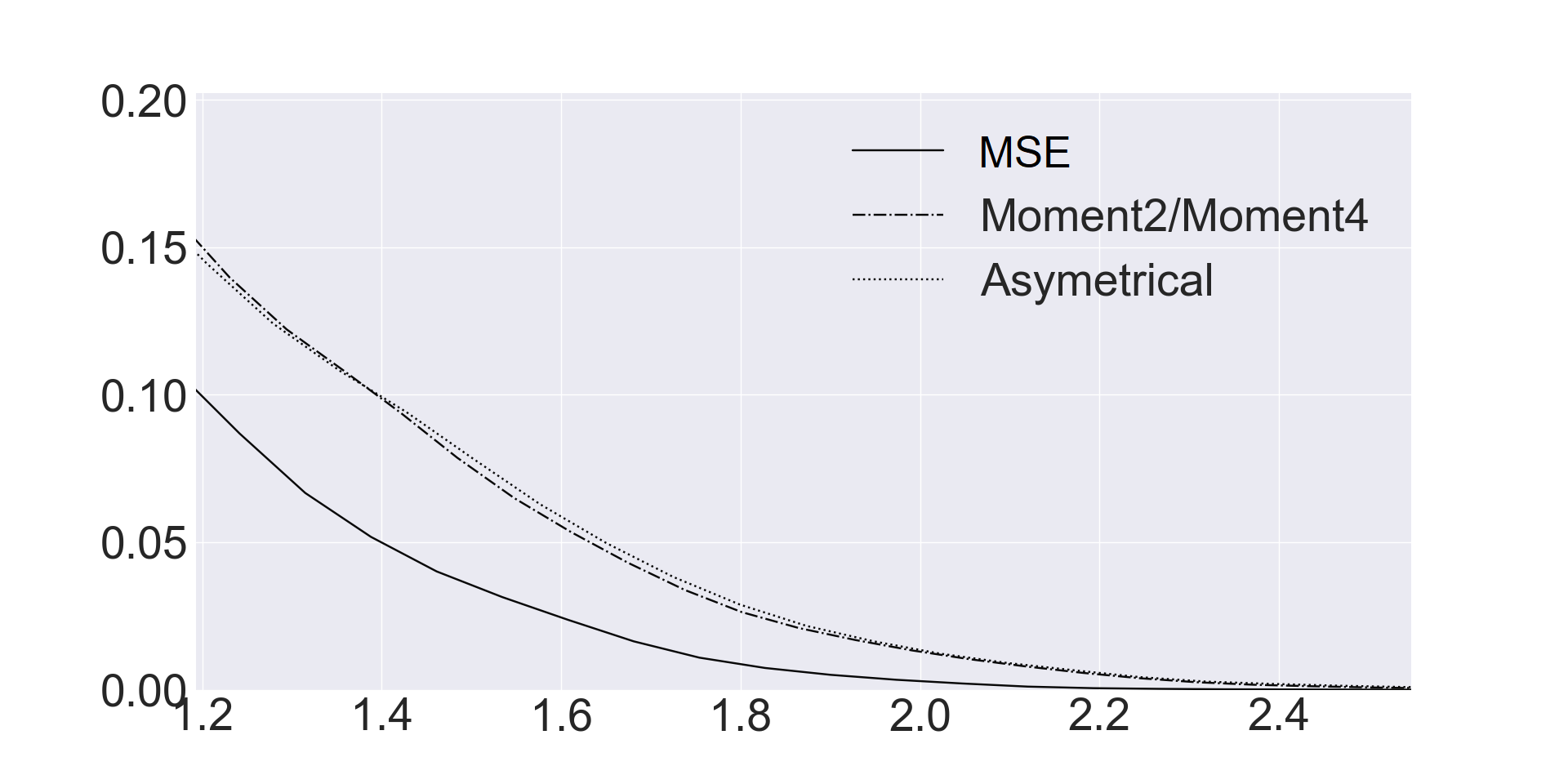}
 \caption*{Zoom on right hand tail}
 \end{minipage}
 \begin{minipage}[b]{0.49\linewidth}
  \centering
 \includegraphics[width=\textwidth]{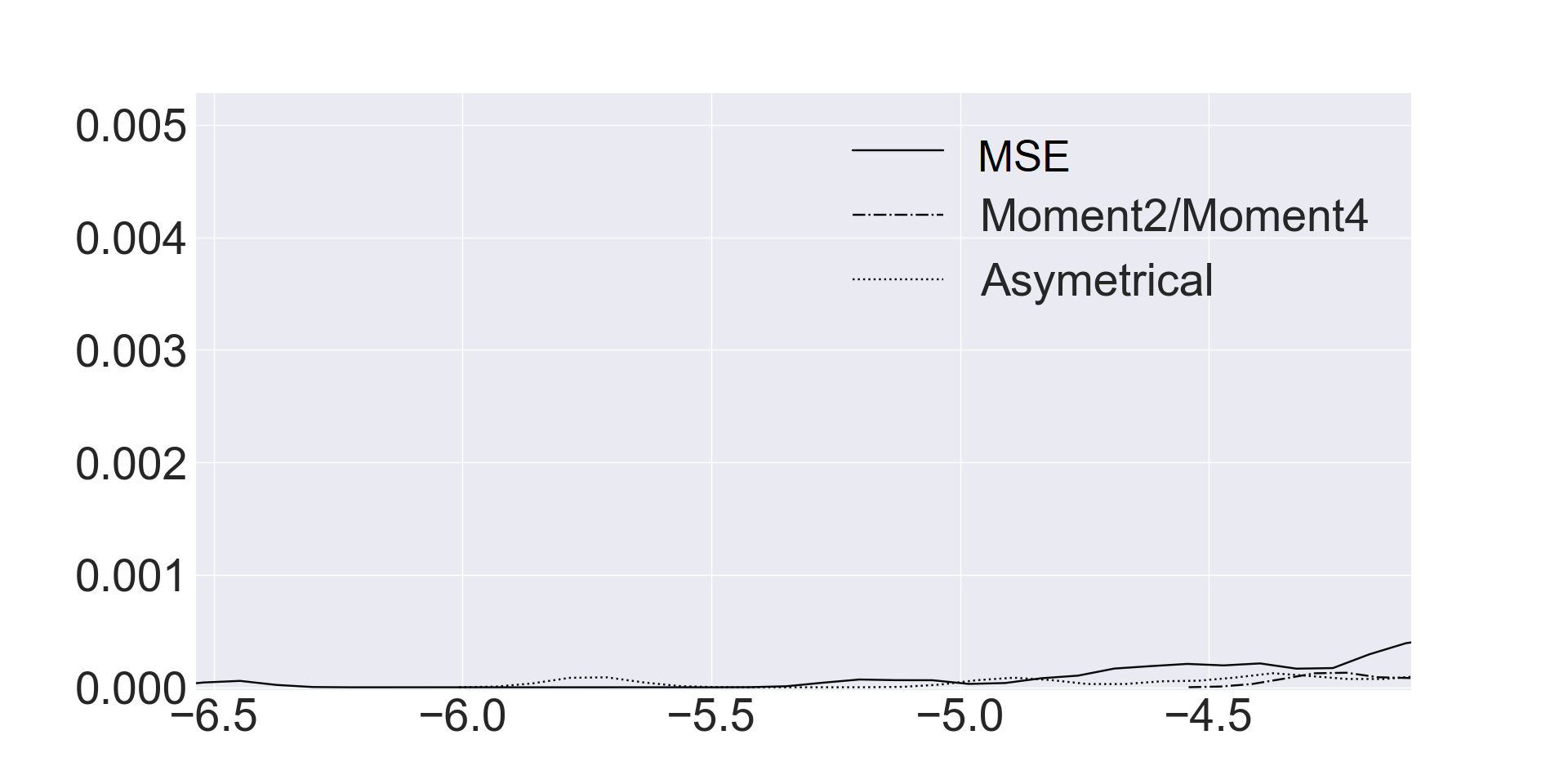}
 \caption*{Zoom on extreme left hand tail}
 \end{minipage}
  \begin{minipage}[b]{0.49\linewidth}
  \centering
 \includegraphics[width=\textwidth]{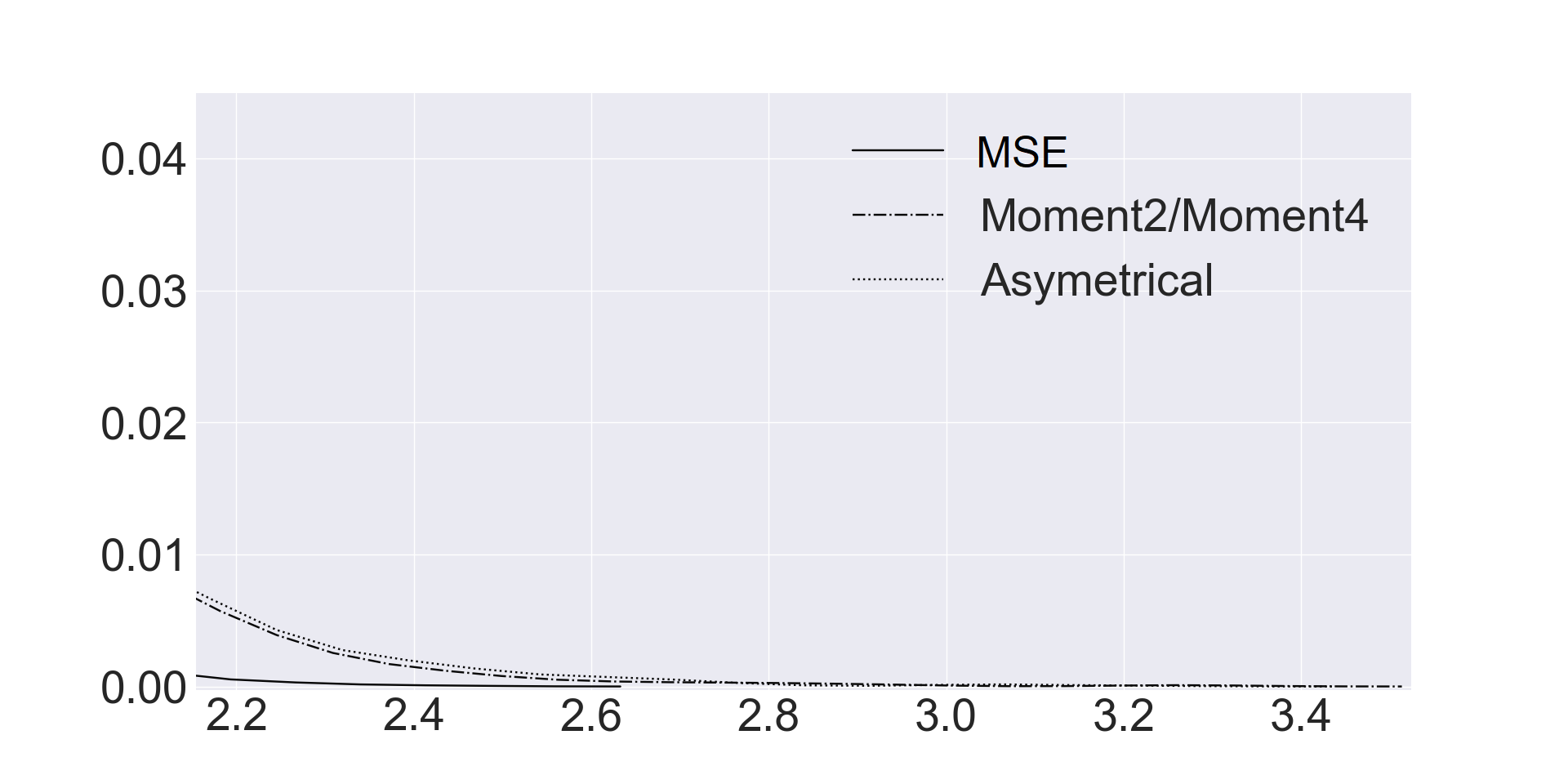}
 \caption*{Zoom on extreme right hand tail}
 \end{minipage}

 \caption{Distribution of the hedged portfolio for Case 3 and different risk criterion - Zoom on the tails}
  \label{fig:riskcase3}
 \end{figure}

One of the advantage of the global neural network approach is its flexibility. There are no particular limitations on the models (Markovian or not, Gaussian or not ...) to use and we can chose different loss functions. In this section, we derive the optimal controls from different losses functions.\\ 
In Figures \ref{fig:riskcase1}, \ref{fig:riskcase2} and \ref{fig:riskcase3}, we plot the distribution\footnote{built with a kernel density estimate method} of the hedged portfolio with the loss functions defined in Equations ( \ref{eq:riskL2}), (\ref{eq:riskAsym}) and (\ref{eq:variancemoment}). In general the non-symmetric losses functions give different shapes for the distributions. On the left hand side, both the asymmetrical loss curve and the Moment 2/Moment 4 loss curve are below the Mean Square loss curve. On the extreme left hand tail represented for example in Figure \ref{fig:riskcase3}, the Mean Square loss function is the only one which is represented: extreme losses are avoided by Moment 2/Moment 4 and asymmetrical loss functions. This is paid on the average (middle of the distribution): there are more minor losses for the two non-symmetrical loss functions. Some of the distribution mass is deported on the right hand side (the gain side). This is an attractive side effect: compared to Mean Squared error, L2/L4 and asymmetrical losses functions tends to favor gains.

\FloatBarrier

\section{Numerical results for option hedging problem with transaction costs}

\label{sec:transactioncosts}

In this section, we investigate the effect of transaction costs when implemented in the global algorithm. We consider that the cost of selling or buying a volume $k$  of $F^i$ is equal to $k.c^i, c^i \geq 0$. As we sell the derivative the terminal wealth of the strategy $X_T$ and associated transaction costs $Y_T$ are equal to: 
\begin{eqnarray*}
X_T & = & p + \sum_{j = 1}^d \sum_{i=1}^{N-1}\Delta^j_{t_i}(S^j_{t_{i+1}}-S^j_{t_i}), \\ 
Y_T &= &\sum_{j = 1}^d \sum_{i=1}^{N-1} |\Delta^j_{t_i} - \Delta^j_{t_{i-1}}| c_j.
\end{eqnarray*}
We use the criterion defined by: 
\begin{eqnarray}
d^{\alpha}(X^\Delta,g(S_T))= (1 - \alpha) \Ebb\left[Y_T \right] + \alpha \sqrt{\Ebb\left[(X_T - g(S_T))^2\right]}, \alpha \in [0,1].
\label{eq:meanvariance}
\end{eqnarray}
This criteria describes a trade-off between risk-limitation and hedging costs. If \(\alpha = 1\), the criterion is equivalent to the variance minimization studied in Section \ref{sec:resultsmeansquare}; if \(\alpha = 0\), we just minimize transaction costs regardless of risks (which corresponds to  doing nothing).\\
\(\alpha \in [0,1]\) is a parametrization of the Pareto frontier of the risk and transaction costs minimization trade-off.
This problem is a portfolio management problem, where $p$ is an input (so not optimized) that we take equal to $\E[g(S_T)]$ in our numerical tests. 
\subsection{Training the Pareto frontier}
Instead of training $N$ versions of the neural network for $N$ values of $\alpha$, we propose to add $\alpha$ to the input variables of the neural network (see Figure \ref{fig:lstm}) and to randomly pick a value of $\alpha$ following a random uniform distribution $\mathcal{U}(0,1)$ at each training iteration. By doing this, we add a dimension to the problem but we obtain the optimal strategy for all $\alpha \in [0,1]$ at once. This goes against traditional algorithms where it is often preferred to evaluate $N$ function defined on $\Rbb^K$ instead of one function defined on $\Rbb^{K+1}$. Getting the whole Pareto frontier is appealing for many reasons as it allows for example to retrieve the $\alpha$ corresponding to an expected transaction cost target budget.\\ 
To obtain the Pareto frontier estimate, we increase the width of the neural network (3 hidden layers of 50 - instead of the 10 previously - neurons for the projection part of the LSTM), and run 100 000 iterations of mini-batch gradient descent while 20 000 were sufficient until now. $\alpha$ is generated from a Sobol quasi random generator. 

\subsection{Numerical results}
\begin{figure}[h!]
\begin{minipage}[b]{0.49\linewidth}
  \centering
 \includegraphics[width=1.\textwidth]{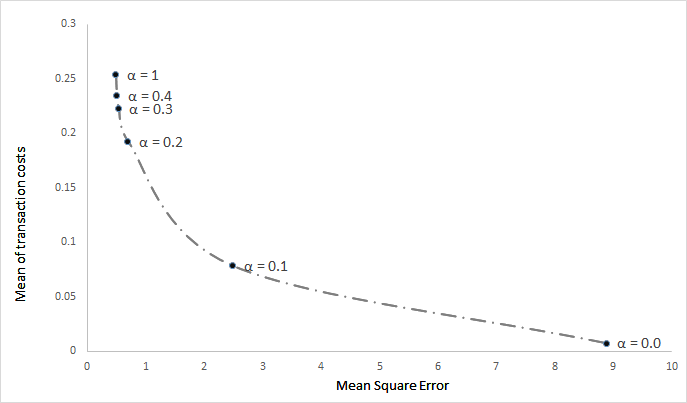}
 \caption*{Case 2. }
 \end{minipage}
\begin{minipage}[b]{0.49\linewidth}
  \centering
 \includegraphics[width=\textwidth]{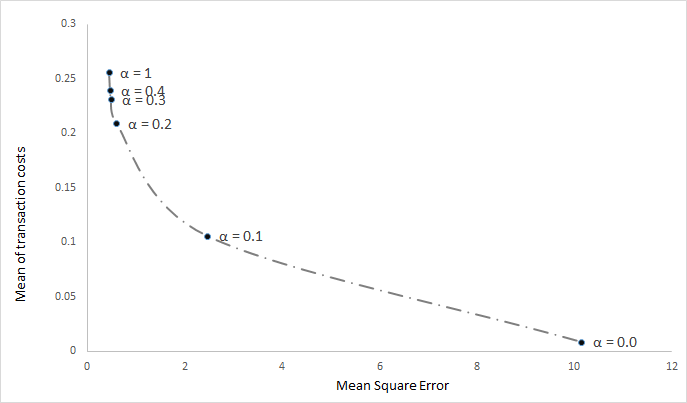}
 \caption*{Case 3. }
 \end{minipage}
 \caption{Spread Option Mean Square VS Mean of transaction costs for various $\alpha$ and transaction cost of 0.02. The dotted line corresponds to a spline interpolation line. }
 \label{fig:frontiere_spread}
 \end{figure}

\begin{figure}[h!]

  \begin{minipage}[b]{0.49\linewidth}
  \centering
 \includegraphics[width=\textwidth]{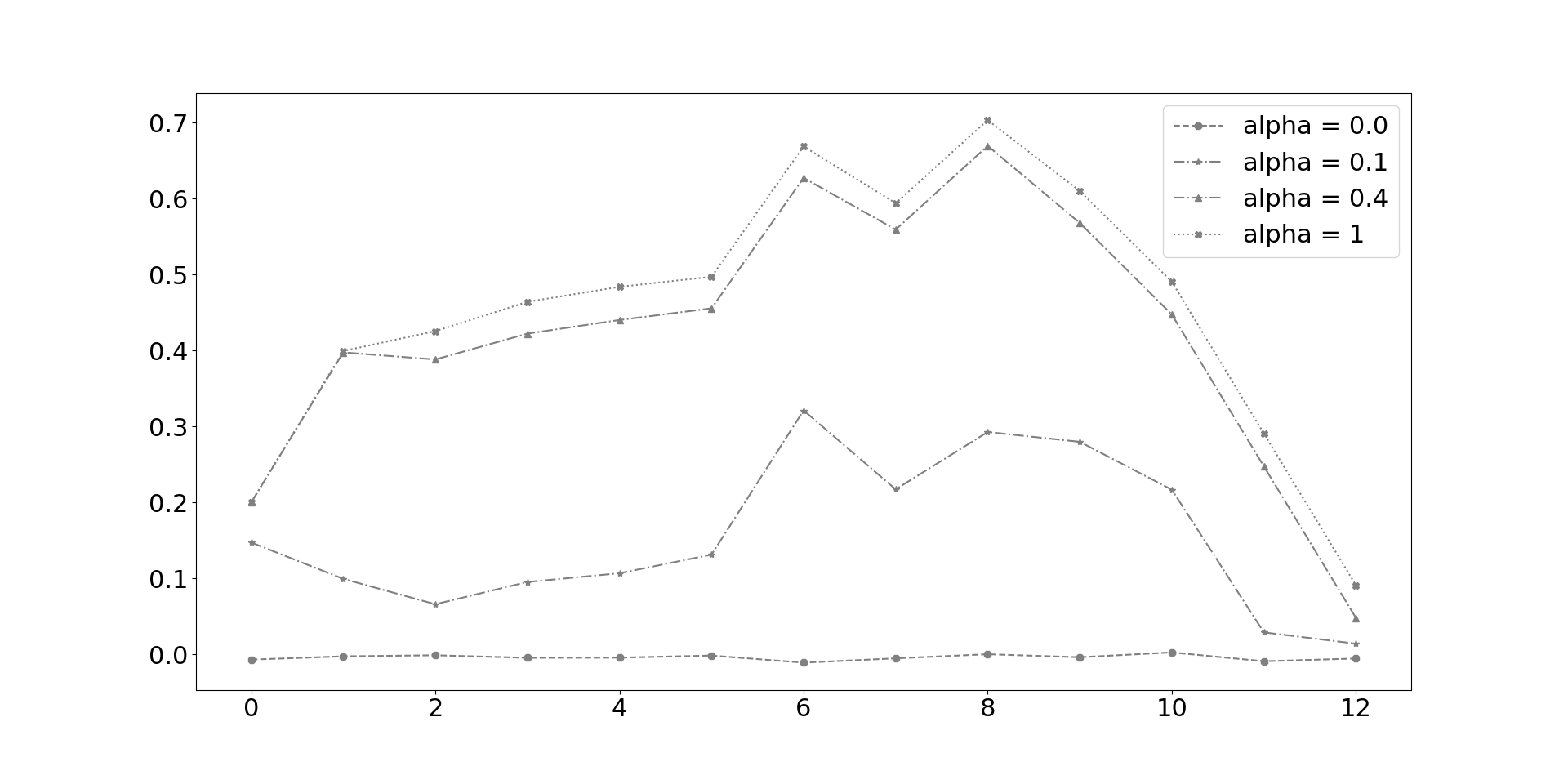}
 \caption*{Sim Nb 1 }
 \end{minipage}
    \begin{minipage}[b]{0.49\linewidth}
  \centering
 \includegraphics[width=\textwidth]{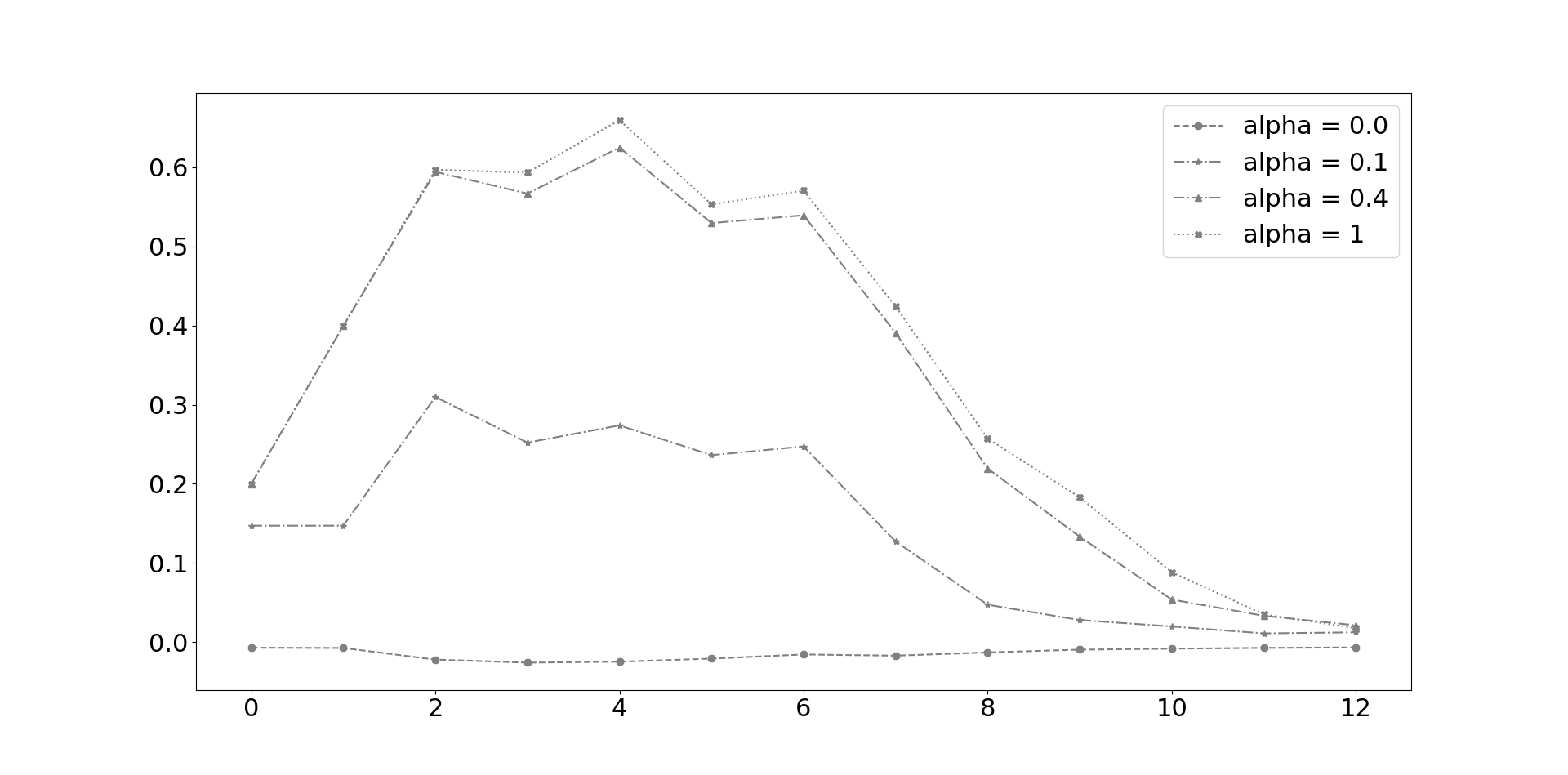}
 \caption*{Sim Nb 2 }
 \end{minipage}
 \caption{Delta for Future 1 and Case 2. and various $\alpha$}
 \label{fig:deltaTCCase2}
 \end{figure}

\begin{figure}[h!]

  \begin{minipage}[b]{0.49\linewidth}
  \centering
 \includegraphics[width=\textwidth]{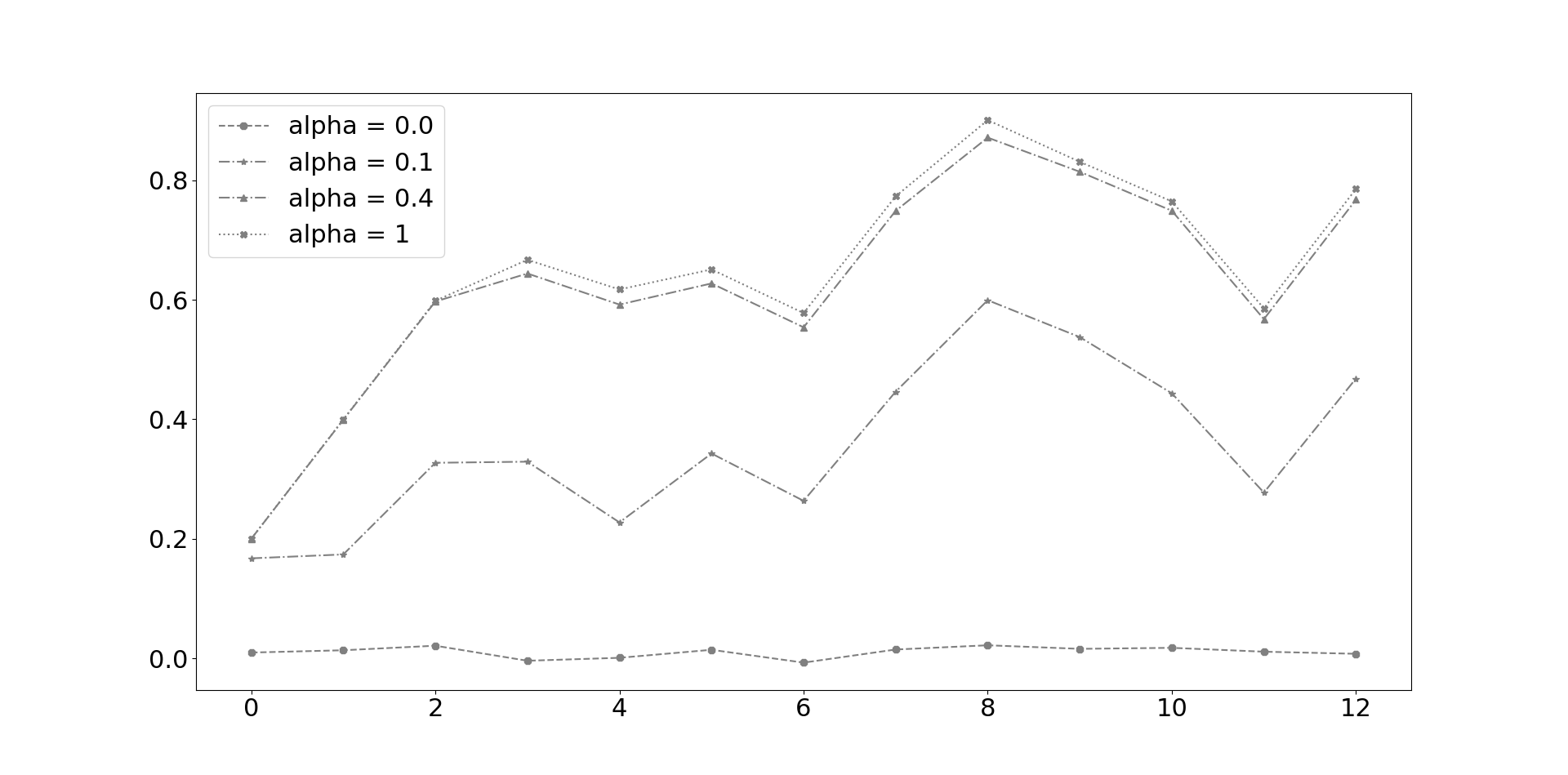}
 \caption*{Sim Nb 1 }
 \end{minipage}
    \begin{minipage}[b]{0.49\linewidth}
  \centering
 \includegraphics[width=\textwidth]{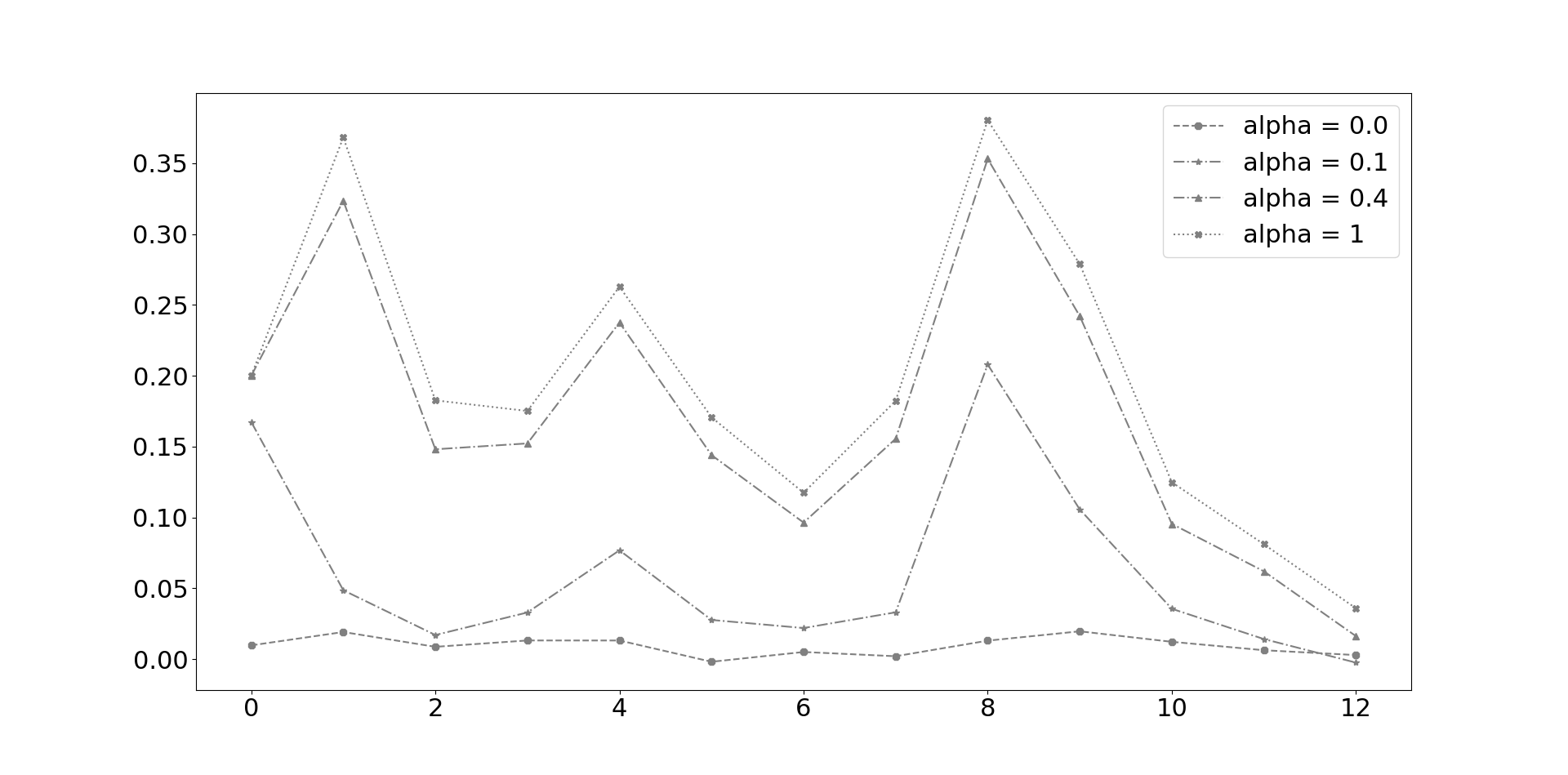}
 \caption*{Sim Nb 2 }
 \end{minipage}
   
 \caption{Delta for Future 1 and Case 3. and various $\alpha$}
 \label{fig:deltaTCCase3}
 \end{figure}

We consider the markets spreads option of Case 2. and Case 3. described in  \ref{sec:spreadoption}. The transaction cost is the same for all tradable risk factors and is set to 0.02 per unit of traded volume. In Figure \ref{fig:frontiere_spread} we plot the resulting average transaction cost and variance of hedged portfolio values for different $\alpha$.
As expected, when $\alpha \sim 1$, the strategy gives similar results to the pure variance minimization of Section \ref{sec:resultsmeansquare}; when $\alpha \sim 0$, we obtain results corresponding to a unhedged portfolio. In Figures \ref{fig:deltaTCCase2} and \ref{fig:deltaTCCase3}, the delta for Case 2 and Case 3 are plotted for some simulations with several $\alpha$'s. For lower $\alpha$ the algorithm prefers to reduce the control amplitude in order to reduce transaction costs. 

\clearpage

\section{Conclusion and perspectives}
Three neural-network-based algorithms (two local algorithms and one global algorithm) dedicated to the hedging of contingent claim are proposed. The three algorithms show good results compared to stochastic-control-based techniques. In particular, the global algorithm is interesting both in terms of execution speed and flexibility. \\ 
The global algorithm is tested with different well known losses function and 
the use of an LSTM architecture in the global algorithm would allow to use some non-markovian underlying models. Moreover, we propose a methodology to draw a Pareto frontier. We apply this methodology to the trade-off between maximizing mean and minimizing variance in the transaction costs case (parameterized by an $\alpha$ combining mean and variance in the objective function). The advantage of getting the whole Pareto frontier is threefold: 
\begin{itemize}
    \item it increases inference speed as we do not need to retrain the algorithm with different parameterization; 
    \item it becomes easy to do a retro-engineering (for example to get which $\alpha$ corresponds to a target transaction costs budget);
    \item it is easier to make sensitivity analysis;
\end{itemize}
The drawback of the global algorithm when compared to stochastic control-based algorithm is the lack of convergence proof. However, the global algorithm allows the treatment of cases that are not attainable by any other techniques.  
\FloatBarrier
\newpage
\bibliography{bibliography}

\end{document}